# ENHANCING RETRIEVAL PERFORMANCE:
# AN ENSEMBLE APPROACH FOR HARD NEGATIVE MINING

HANSA MEGHWANI

Thesis Report

JUNE 2024

# DEDICATION

This thesis is dedicated with gratitude and appreciation to my family and friend Kumar Mayank who have provided unwavering support and encouragement throughout my academic journey. Their collective wisdom and encouragement have been a bedrock of my determination and success.

**ACKNOWLEDGEMENTS**

First and foremost, I extend my deepest gratitude to my thesis supervisor, Mrs. Harika Vuppala, for her invaluable guidance and encouragement throughout this thesis. Without her direction, completing this work would have been unimaginable. I am also grateful to my fellow research students, the instructors from LJMU, and the upGrad support staff for their assistance in making this academic journey smoother.

I would also like to express my gratitude to Oracle India for providing the dataset and supplying the essential computing resources required for this research. Additionally, I must thank my family, friends, and colleagues, who have consistently supported me throughout this journey. Their encouragement was instrumental in my decision to pursue the master's program and successfully complete this thesis.

## ABSTRACT

Ranking consistently emerges as a primary focus in information retrieval research. Retrieval and ranking models serve as the foundation for numerous applications, including web search, open domain QA, enterprise domain QA, and text-based recommender systems. Typically, these models undergo training on triplets consisting of binary relevance assignments, comprising one positive and one negative passage. However, their utilization involves a context where a significantly more nuanced understanding of relevance is necessary, especially when re-ranking a large pool of potentially relevant passages. Although collecting positive examples through user feedback like impressions or clicks is straightforward, identifying suitable negative pairs from a vast pool of possibly millions or even billions of documents possess a greater challenge. Generating a substantial number of negative pairs is often necessary to maintain the high quality of the model. Several approaches have been suggested in literature to tackle the issue of selecting suitable negative pairs from an extensive corpus. This study focuses on explaining the crucial role of hard negatives in the training process of cross-encoder models, specifically aiming to explain the performance gains observed with hard negative sampling compared to random sampling. We have developed a robust hard negative mining technique for efficient training of cross-encoder re-rank models on an enterprise dataset which has domain specific context. We provide a novel perspective to enhance retrieval models, ultimately influencing the performance of advanced LLM systems like Retrieval-Augmented Generation (RAG) and Reasoning and Action Agents (ReAct). The proposed approach demonstrates that learning both similarity and dissimilarity simultaneously with cross-encoders improves performance of retrieval systems.

# TABLE OF CONTENTS

## LIST OF FIGURES

## LIST OF TABLES

## LIST OF ABBREVIATIONS

| | |
|---|---|
| RAG……………………………………. | Retrieval Augmented Generation |
| ReAcT……………………………….. | Reasoning and Action Agent |
| GenAI………………………………... | Generative Artificial Intelligence |
| WAF…………………………………. | Web Application Firewall |
| VCN…………………………………. | Virtual Cloud Network |
| BERT………………………………... | Bi-directional Encoder Representation from Transformers |
| RoBERTa……………………………. | Robustly Optimized BERT Pre-Training Approach |
| IR…………………………………….. | Information Retrieval |
| LLM…………………………………. | Large Langage Model |
| NLP………………………………….. | Natural Language Processing |
| DPR………………………………….. | Dense Passage Retrieval |
| BM25………………………………... | Best Match 25 |
| ADORE……………………………… | Algorithm for Directly Optimizing Ranking performance |
| STAR………………………………... | Stable Training Algorithm for dense Retrieval |
| ANCE………………………………... | Approximate nearest neighbour Negative Contrastive Estimation |

# Chapter 1 INTRODUCTION

## 1.1 Background of The Study

Retrieval models, unlike generative models, fetch real information from sources, with search engines indicating the source of each retrieved item (Sanderson and Croft, 2012). This underscores the continued importance of information retrieval (IR), even in the presence of generative LLMs, particularly in contexts where reliability is crucial. After the launch of BERT (Devlin, Chang, Lee, Google, et al., 2019) , a straightforward retrieve then re-rank strategy became popular in January 2019 as a successful means of leveraging pre-trained transformers for passage retrieval (Nogueira and Cho, 2019a) . This model, known as monoBERT, marks the initial manifestation of what later evolved into cross-encoders for retrieval, a category encompassing reranking models such as MaxP (Dai and Callan, 2019), CEDR (MacAvaney et al., 2019a), Birch (Akkalyoncu Yilmaz et al., 2019) PARADE (Li et al., 2020), and numerous others. A proficient ranking algorithm holds potential advantages for numerous downstream tasks within information retrieval research (Han et al., 2020). Conventional algorithms like BM25 (Robertson and Walker, 1994) heavily rely on term-matching metrics, limiting their efficacy to scenarios where queries and documents share identical terms. This inherent drawback leads to performance degradation when faced with semantic differences despite identical meanings, a phenomenon known as the vocabulary mismatch problem. To enhance the understanding of users' search intentions and retrieve pertinent items, it's anticipated that ranking algorithms would engage in semantic matching between queries and documents. Driven by advancements in deep learning, particularly techniques for capturing meaning (representation learning), researchers are increasingly using Dense Retrieval (DR) models to tackle the challenge of semantic similarity (Guu et al., 2020; Karpukhin et al., 2020; Luan et al., 2020). DR excels at capturing the semantic essence of queries and documents by converting them into low-dimensional embeddings. This facilitates efficient document indexing and similarity search, leading to effective online ranking. Studies have shown promising results for DR models in various information retrieval tasks. (Guu et al., 2020; Qu et al., 2020)

While past research using diverse training strategies for DR models has shown encouraging results, inconsistencies and even contradictions arise when comparing their findings. For example, the superiority of training with “hard negatives” (highly similar yet irrelevant documents) over random negatives remains an open question. Additionally, many effective training methods suffer from inefficiency, making them impractical for large-scale deployments. Despite promising results, DR faces key challenges, and we are trying to investigate one of the challenges related to hard negatives mining on a custom data which has specific industry context.

## 1.2 Problem Statement

For the integration of Generative AI, every organization will require a custom retrieval system built upon their private datasets. Retrieval systems, when not trained on enterprise domain datasets, face significant challenges in delivering accurate and relevant results within organizational contexts. One of the primary issues is a lack of familiarity with the specific terminology, jargon, and nuances prevalent in the enterprise domain.

The major issues while training the retrieval models -

- Hard negatives, which are non-relevant passages that closely resemble positive examples, play a crucial role in refining the model's understanding.
- Providing both positive (relevant) and negative (irrelevant) examples is important. Negative examples, especially hard negatives, challenge the model to distinguish between relevant and irrelevant content effectively.
- When dealing with enterprise-specific datasets, the inclusion of hard negatives is paramount.
- Without exposure to hard negatives, the model might struggle to differentiate between similar passages, leading to inaccurate responses and compromised decision-making processes within organizations.

## 1.3 Aim and Objectives

The principal aim of this research is to propose a hard negative mining strategy which helps enterprises to fine-tune state of the art ranking/retrieval models on their private dataset. These datasets are unique to each enterprise and may contain specialized terminologies and domain-specific jargon. So, by incorporating hard negatives in the training process, retrieval models can be honed to navigate the complexities of enterprise data, ensuring precise and contextually relevant results tailored to the specific needs of the business.

The research objectives are derived from the purpose of this study, which are outlined as follows:

- To suggest a robust hard negative mining strategy for domain specific private data.
- To utilize the hard negatives for fine-tuning of cross-encoder model on domain specific data.
- To investigate the impact of using hard negatives in training of ranking models.

## 1.4 Significance of the Study

Large Language Models (LLMs) excel at answering questions based on the knowledge they were trained on. However, their training data typically does not include recent information and specific private information stored in platforms like a company's Confluence, Google Drive, or SharePoint. For the integration of Generative AI within enterprise settings, every organization will require a custom retrieval system built upon their private datasets. Retrieval augmented systems, when not trained on enterprise domain datasets, face significant challenges in delivering accurate and relevant results within organizational contexts. One of the primary issues is a lack of familiarity with the specific terminology, jargon, and nuances prevalent in the enterprise domain. Importance of training retrieval/ ranking models with hard negatives cannot be overstated, especially within enterprise domains. This study bridges the gap between pre-trained ranking models and domain-specific data. The study also proposes a data augmentation technique to include hard negatives for given query, document pair. This study will investigate that the triplet objective function can be utilized by researchers across various NLP applications. Additionally, it aims to introduce a modified evaluation criterion suitable for use by researchers evaluating ranking models for internal dataset.

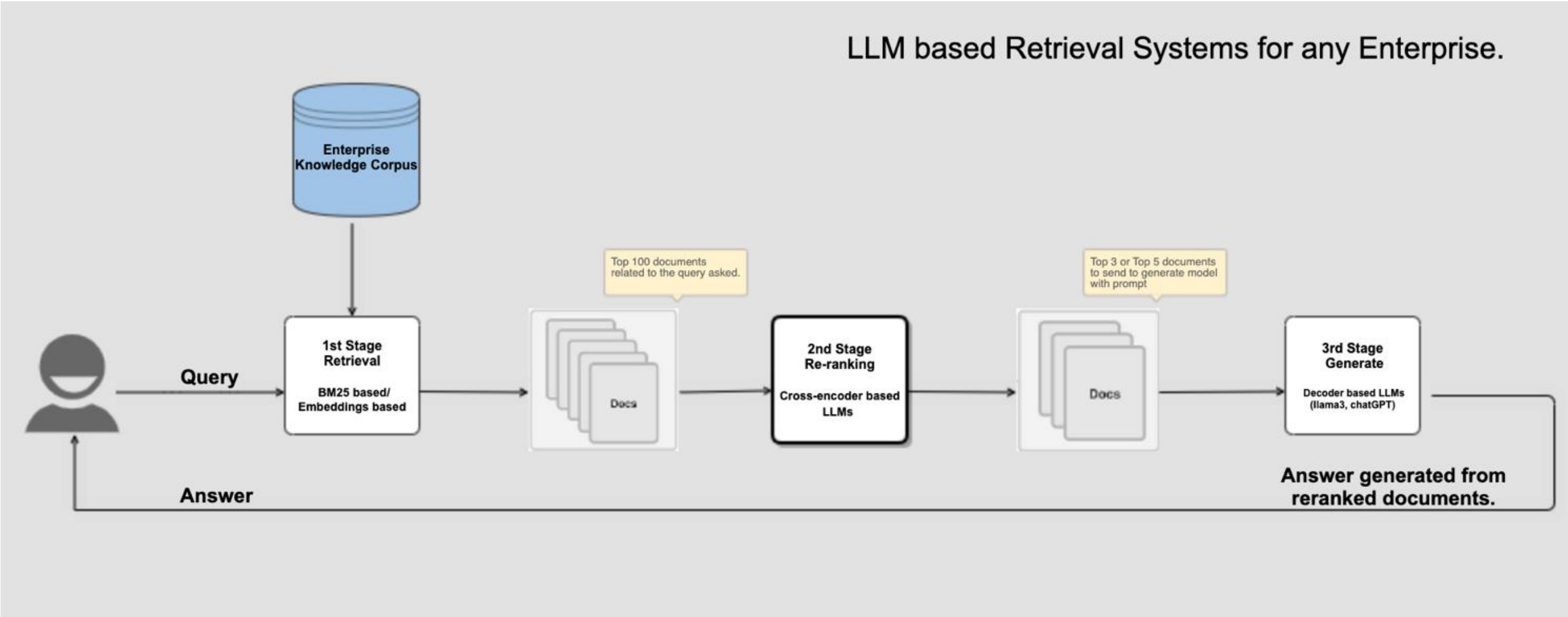

**Figure 1.1** Significance of research study in LLM based retrieval systems for any enterprise.

Figure 1.1 provides on overview of the system; this retrieval system is the need of hour for every enterprise to leverage Generative AI or LLM application for their use case. The proposed research study is focused on improving the performance of $2^{nd}$ stage which is reranking of documents before passing it to LLM generate model.

## 1.5 Scope of the Study

The scope of the study is to develop hard negative mining strategy using ensemble of embeddings. Pre-trained embedding models used here will be state of the art models like SFR-Embedding-Mistral, Jina AI, BERT etc. And to analyse impact of fine-tuning cross encoder models with mined hard negatives on the private enterprise dataset. Given resource limitations, this study will not explore training embedding models from scratch using large-scale text corpora. We will instead leverage pre-trained embeddings and adapt the size and complexity of candidate models to fit within available resources. Training models approaching the scale of leading-edge architectures is not feasible for this study.

### 1.6 Structure of the Study

The structure of this research study is outlined as follows. Chapter 1 covers details about the research proposal like background of the study, problem statement, scope and significance of doing this research.

Chapter 2 provides an in-depth examination of the literature on hard negative mining techniques and cross-encoder training for document and passage ranking tasks. Sections 2.2 and 2.3 detail the literature on hard negatives and various negative sampling techniques. Section 2.4 discusses the literature related to retrieval and ranking models, specifically focusing on bi-encoders and cross-encoders. Section 2.5 outlines advanced techniques for generating synthetic data for model training. Section 2.6 addresses downstream tasks like document ranking and the associated open-source datasets. Finally, in section 2.7 reviews various loss functions used for these downstream tasks. This study covers intersections among hard negative mining, ensembles, and cross-encoder re-rankers, mandating a comprehensive review of the applicable literature.

Chapter 3 provides a comprehensive explanation of the methods used in this study to improve the document ranking task by augmenting existing data with hard negative mining. Section 3.2 presents the proposed flow diagram. Section 3.3 details the dataset description and preprocessing steps taken to understand the data's characteristics. Following that, Section 3.4 discusses how data augmentation is applied using an embeddings ensemble and describes the model architecture proposed for mining hard negatives and re-ranking documents. The chapter concludes in Section 3.5, where it outlines the evaluation metrics that will be used to assess the model after training.

Chapter 4 outlines the implementation of the approaches mentioned in the research methodology. It includes data exploration, analysis, and experimentation aimed at uncovering new insights and observations about the enterprise dataset. Section 4.2 provides detailed information about the implementation of embedding models, the ensemble technique used to select hard negatives, and the data preparation and hyperparameters for cross-encoder training. Section 4.3 examines the distribution of word and token counts in queries and documents. Section 4.4 details the hardware and software requirements used in this research.

Chapter 5 is about results for the downstream task of document ranking on an enterprise dataset used in retrieval-based industry applications like Retrieval Augmented Generation (RAG). Section 5.2 will cover the results of using embedding ensembles and clustering techniques to select hard negatives. In Section 5.3, provides outcomes of training the cross-encoder based re-ranker model on the enterprise dataset. Section 5.4 describes few limitations of this study.

Chapter 6 summarizes the thesis by outlining the contributions and conclusions. Based on the experiments conducted, we offer basic recommendations for improvements to aid future research in this field.

# Chapter 2 LITERATURE SURVEY

## 2.1 Introduction

This section of the literature review conducts an in-depth examination of hard negative mining techniques and cross-encoder training for document and passage ranking tasks. We will initially explore the task of hard negative mining and the ways in which it has been categorized in the literature. Subsequently, we will delve into the research concerning cross-encoder models, embedding models, and several synthetic document generation methods. This study intersects hard negative mining, ensembles, and cross-encoder re-rankers, prompting a thorough review of the related literature as well.

## 2.2 Hard Negative

Previous studies highlight the critical role of negative example selection in optimizing the training process of cross-encoder models. (Karpukhin et al., 2020) conducted a study to assess the efficacy of different negative sampling strategies in training. They compared the performance of BM25 (Robertson and Walker, 1994) negatives, random negatives, and in-batch negatives. Their findings suggest that combining BM25 and in-batch negatives leads to the most favourable outcomes in terms of training effectiveness. (Xiong et al., 2020) theoretically demonstrate that employing local negatives is suboptimal for dense retrieval learning. Subsequently, they suggest a method for generating global negatives by utilizing the existing dense retrieval model concurrently during training. This approach necessitates periodic re-indexing of the corpus and retrieval. (Qu et al., 2020) additionally suggest a method for generating hard negatives by utilizing the existing dense retrieval model, albeit after the completion of training rather than dynamically during training. However, their study reveals that employing these hard negatives alone post-training could potentially hinder the training process. They note that the effectiveness of these hard negatives is improved when filtered based on an independently trained cross-encoder model. (Zhan et al., 2021) discover that the instability induced by hard negatives can be mitigated by incorporating random negatives into the training process. Furthermore, they adopt a strategy akin to the ANCE (Xiong et al., 2020) method for periodically re-generating hard negatives, with the modification of updating only the query encoder to reduce re-indexing overhead. These findings underscore the significance of hard negatives in training and demonstrate varying levels of effectiveness across different approaches. Apart from the aforementioned research, which primarily delves into hard negative training strategies for fine-tuning DPR-like bi-encoders, other studies have made comparable observations regarding diverse methodologies that benefit bi-encoders.  highlight the continued importance of hard negatives even when the model undergoes additional pretraining aimed at enhancing the representation of the [CLS] token, referred to as Condenser in their study. The significance of hard negative mining has also been demonstrated in scenarios where knowledge distillation is implemented on biencoders (Gao and Callan, 2021). (Hofstätter et al., 2021) In contrast to the numerous studies investigating the benefits of hard negatives for bi-encoders, only one study (Gao and Callan, n.d.) successfully integrates hard negatives into cross-encoder training. They introduce the Localized Contrastive Estimation (LCE) loss to illustrate its efficacy, demonstrating that integrating this loss function with harder negatives substantially enhances reranking performance, particularly when the distribution of training instances aligns with the results retrieved by initial-stage retrievers. (Pradeep et al., 2022)

## 2.3 Negative Sampling Strategies

Negative sampling is a crucial aspect of training models for various tasks, including information retrieval. It involves selecting examples that are irrelevant to the query, allowing the model to learn the distinction between positive and negative examples. Different negative sampling techniques have varying effectiveness. Random negatives might be easy to generate but may not provide the most informative contrast to positive examples. BM25 negatives, leveraging a retrieval model like BM25, could be a more focused approach but might introduce bias from the BM25 model itself.

### 2.3.1 BM25 Negatives

This stands for "Best Match 25" and is a popular retrieval model used in information retrieval systems. It ranks documents based on their relevance to a query, considering factors like term frequency (how often a term appears in a document) and document frequency (how many documents contain the term).

**Table 2.1** Advantages and Disadvantages of BM25 Negatives

| Advantages | Disadvantages |
|---|---|
| **Targeted Selection:** Compared to random negatives, BM25 negatives offer a more focused approach. The negatives are specifically chosen from the same document pool as the potential answers, ensuring they share some context with the query. | **Bias from BM25:** BM25 itself might have limitations or biases. If the training data used for BM25 is skewed, it could introduce those biases into the selection of negative examples. |
| **Leveraging Existing Knowledge:** BM25 is a well-established retrieval model, so using its scoring for negative selection can be | **Not Perfect Relevance Indicator:** While BM25 scores provide a good estimate of relevance, they might not be perfect. There could be passages |
| computationally efficient and leverage existing knowledge about relevance. | with low scores that are still somewhat relevant to the query. |

### 2.3.2 In-batch Negatives

A batch typically contains multiple queries and their corresponding candidate answer passages. Instead of searching a larger dataset for irrelevant examples, in-batch negatives leverage the passages already present within the current batch. Passages that are unlikely to be relevant to the specific query at hand are chosen as negative examples.

**Table 2.2** Advantages and Disadvantages of In-batch Negatives

| Advantages | Disadvantages |
|---|---|
| **Computational Efficiency:** Since negatives are chosen from within the current batch, it reduces the need to search a larger dataset for irrelevant examples, potentially saving computational resources. | **Limited Scope:** Negatives are restricted to the current batch, which might limit the diversity of irrelevant examples the model sees during training. |
| **Context-aware Selection:** Negatives are chosen from the same pool of passages as potential answers, ensuring they share some context with the query, which can be beneficial for training. | **Potential for Bias:** Depending on how negatives are selected within the batch, there might be a bias towards certain types of irrelevant passages. |

### 2.3.3 Random Negatives

- The model randomly selects documents or passages from the entire dataset, regardless of their content or relevance to the current query.
- Randomly selecting negative examples during training can lead to suboptimal performance. These randomly chosen documents may not provide informative gradients for model updates.

**Table 2.3** Advantages and Disadvantages of Random Negatives

| Advantages | Disadvantages |
|---|---|
| **Simplicity:** This approach is very easy to implement. You don't need any complex | **Low Informativeness:** Randomly chosen negatives might not be very informative for training. |
| algorithms or pre-processing steps to choose negative examples. | |
| **Computational Efficiency:** Since random selection is involved, it's computationally inexpensive compared to techniques that require scoring or ranking documents. | They might be completely unrelated to the task at hand, making it harder for the model to learn the distinction between relevant and irrelevant information. |

### 2.3.4 Comparing random negatives, BM25 negatives and in-batch negatives

(Karpukhin et al., 2020) investigates different negative sampling strategies for training dense retrieval models.

- (Karpukhin et al., 2020) compare the effectiveness of random negatives with other methods and find that random selection alone is not ideal. (Pradeep et al., 2022) The authors propose a static hard negative mining strategy called Passage-based BM25 (Nguyen et al., n.d.). Instead of selecting negatives based on the query, PassageBM25 selects negative passages based on their similarity to the positive passage.
- Empirical studies on a Vietnamese question-answering dataset (ZAC2022) demonstrate that PassageBM25 outperforms both vanilla BM25 and dense retrievers trained with conventional query-based BM25 in terms of top-k retrieval accuracy. (Nguyen et al., n.d.)
- (Xiong et al., 2020) theoretically prove that local negatives (such as in-batch negatives) are suboptimal for dense retrieval learning.

### 2.3.5 Combining BM25 and In-Batch Negatives

- (Karpukhin et al., 2020) find that a mixture of BM25 and in-batch negatives yields optimal results. By combining these two types of negatives, the model benefits from both static and dynamic negative mining strategies.
- Additionally, RocketQA (Qu et al., 2020) proposes an optimal pipeline for training bi-encoder models, which involves increasing batch size through in-batch and cross-batch training and denoising false negative samples.

In summary, the choice of negative sampling strategy significantly impacts the performance of dense retrieval models. While random negatives are suboptimal, combining BM25 and in-batch negatives provides an effective approach to enhance retrieval accuracy (Pradeep et al., 2022)

### 2.3.6 Static and Dynamic Hard Negatives (STAR & ADORE)

(Zhan et al., 2021) theoretically analyse the impact of negative sampling strategies on DR model performance. They shed light on why hard negative sampling (selecting challenging negatives) outperforms random sampling. Additionally, they identify potential risks associated with static hard negative sampling.

The paper introduces two novel training strategies:

- Stable Training Algorithm for dense Retrieval (STAR):
  - STAR combines two types of negatives during training random negatives alongside static hard negatives.
  - Random negatives help stabilize the training process and mitigate risks associated with static hard negatives.
  - Additionally, STAR reuses document embeddings within the same batch to enhance efficiency.
  - By introducing random negatives alongside static hard negatives, STAR aims to achieve better stability during training.
- Query-side training Algorithm for Directly Optimizing Ranking performance (ADORE):
  - ADORE dynamically selects hard negatives during training, rather than relying on a fixed set.
  - ADORE replaces static hard negative sampling with a dynamic method that directly optimizes ranking performance.
  - By dynamically selecting negatives, ADORE aims to improve overall retrieval accuracy.

The proposed strategies are evaluated on two publicly available retrieval benchmark datasets. Both STAR and ADORE demonstrate significant improvements over existing competitive baselines. Combining both strategies yields the best overall performance.

Drawbacks:

- **Trade-off Between Stability and Effectiveness:** STAR balances stability (due to random negatives) with effectiveness (due to hard negatives). However, finding the right balance can be challenging. Striking the optimal trade-off remains an open question.
- **Dynamic Sampling Complexity:** ADORE dynamically selects hard negatives during training, which introduces additional complexity. The process of dynamically identifying suitable negatives requires careful design and efficient implementation.

- **Risk of Overfitting:** Dynamic negative sampling may inadvertently lead to overfitting if the model becomes too specialized to the training data. ADORE needs to strike a balance between optimizing ranking performance and avoiding overfitting.

## 2.4 Retrieval and Reranking models

In this section of the research study, we will discuss about various retrieval and re-ranking models. These are mostly encoder type models. Cross encoder returns a classification score between 0 to 1 which signifies the similarity between the input documents. So, the cross-encoder is used for re-ranking the documents, passages and other texts. On the other hand, bi-encoder model can return individual embeddings of the documents or texts as an output and hence these are used to get embeddings of documents. In the section 2.4.1 we will summarize the earlier research conducted in cross-encoder models and in the section 2.4.2 we will summarize the research work done in bi-encoders models.

### 2.4.1 Cross encoder

The initial cross-encoder for reranking, monoBERT (Nogueira and Cho, 2019a), swiftly emerged shortly after the introduction of BERT (Devlin et al., 2019a) It adhered to the approach advocated by the BERT authors for processing (query, document) input pairs and exhibited substantial advancements in effectiveness on datasets such as TREC CAR (Craswell et al., n.d.; Dietz et al., n.d.) and MS MARCO passage ranking (Bajaj et al., 2016). While the vanilla version of monoBERT (Nogueira et al., 2019) demonstrated significant enhancements in the passage retrieval task, its design did not accommodate the processing of long input sequences necessary for document retrieval. Many subsequent BERT-based cross-encoder studies (MacAvaney et al., 2019a; Li et al., 2020) aimed to tackle this challenge by either conducting multiple inferences on various segments of the document or implementing additional architectural modifications atop BERT to enhance the processing of longer document texts.

### 2.4.2 Bi encoder

Biencoders encode each sentence separately and map them to a common embedding space, where the distances between them can be measured. They are fast, scalable, and efficient for inference. This makes them suitable for tasks that require efficient comparison in a vector space, such as information retrieval or semantic search. However, they are generally less accurate than cross-encoders. The breakthroughs in DPR (Karpukhin et al., 2020) and ANCE (Xiong et al., 2020) have sparked a revival of biencoders in the BERT era. The aim of a bi-encoder is to learn a transformer-based mapping from queries and documents into dense fixed-width vectors, where the inner product between the query vector and the relevant document vector is maximized. Significant efforts have been made to understand and improve this mapping (Lindgren et al., n.d.; Hofstätter et al., 2020; Gao et al., 2021)

(Lin et al., 2020) have developed a training method for a biencoder that employs real-time knowledge distillation from a ColBERT (Khattab and Zaharia, 2020) teacher model. This model calculates soft labels for negatives within the batch. The KL-divergence is used to capture the difference between the score distributions of the student and teacher models for all instances in the batch. It has been demonstrated that the addition of this loss to the conventional cross entropy loss over relevance labels leads to improved scores.

## 2.5 Generation of Synthetic Document for Training Cross Encoder

(Askari et al., 2023; Agarwal et al., 2024) investigates the effectiveness of using large language models (LLMs) like ChatGPT to generate synthetic documents for training cross-encoder re-rankers. Cross-encoder re-rankers are a technique in information retrieval that improves the relevance of search results by re-ordering them based on semantic similarity to the query. The study compares the performance of re-rankers trained on synthetic data generated by ChatGPT with those trained on human-written data. The findings suggest that LLM-generated documents can be a viable alternative, particularly in domains with limited labelled data. The research also introduces a new dataset, ChatGPT-RetrievalQA, which is based on HC3 dataset (Guo et al., 2023) for evaluating the effectiveness of LLM-generated data in this context. Overall, the paper contributes to the exploration of techniques for improving information retrieval systems using synthetic data. And few other research (Agarwal and Pachauri, 2023; Agarwal et al., 2023) explores pseudo labelling and graph data structure for data augmentation.

### 2.5.1 Research Question

- The study aims to compare cross-encoder re-ranking models trained on ChatGPT responses with those trained on human responses.
- Specifically, they investigate whether LLM-generated data can augment training data, especially in domains with limited labelled data.

### 2.5.2 Findings

- Zero-Shot Re-Ranking: Cross-encoder re-rankers trained on ChatGPT responses are statistically significantly more effective as zero-shot re-rankers than those trained on human responses.

- Supervised Setting: In a supervised setting, human-trained re-rankers outperform LLM-trained re-rankers.
- Generative LLMs show promise in generating training data for neural retrieval models.

### 2.5.3 Drawbacks

- Control and Transparency: It can be challenging to precisely control the content and style of documents generated by large language models. This lack of control over the training data might make it difficult to optimize the re-rankers for specific tasks.
- Factual Incorrectness: Further research is needed to explore the impact of factually incorrect information in the generated responses and to validate these findings with open-source LLMs.
- Limited Domain Transferability: ChatGPT might be less effective in generating documents for specialized domains where its training data is limited. The paper mentions limitations in ChatGPT for cross-linguistic tasks, which could also be a disadvantage.
- Cost and Computational Resources: Training large language models like ChatGPT requires significant computational resources. This could be a barrier for wider adoption of the proposed method. Also, for domain specific human annotated dataset, expert human annotators are required, which is very expensive and tedious task.

## 2.6 Document Ranking Datasets

In the document ranking task, the goal is to rank documents from a large corpus based on their relevance to a query.

There are two subtasks within this task:

- Full ranking (retrieval): Rank documents comprehensively based on relevance.
- Top-100 reranking: Further refine the ranking by considering the top 100 retrieved documents.

### 2.6.1 MS MARCO

MS MARCO was introduced with a paper released at NIPS 2016. (Bajaj et al., 2016) It encompasses various datasets, each focusing on different aspects of machine reading comprehension, question answering, and ranking tasks. The initial dataset included 100,000 real Bing questions paired with human-generated answers. Subsequent releases expanded to include more diverse tasks and larger question sets.

The corpus for the document ranking task contains 3.2 million documents. The training set consists of 367,013 queries with associated relevance labels for documents. Researchers use this dataset to evaluate and benchmark various ranking models. Transformer-based models, such as BERT (Devlin et al., 2019a) have also been used for re-ranking documents in this task.

### 2.6.2 TREC (Text REtrieval Conference) Dataset

The TREC Deep Learning (DL) Track (Dietz et al., n.d.; Craswell et al., 2021) focuses on information retrieval in a large training data regime. Specifically, it provides large-scale training data with tens of thousands or even hundreds of thousands of queries, each having at least one positive label. This mirrors real-world scenarios, including training based on click logs and labels from shallow pools. The DL Track aims to investigate which methods perform best under these conditions. Key questions include whether methods effective on small data also work on large data, how much performance improves with more training data, and how external data and weak supervision enhance results. In summary, the track offers valuable resources for advancing deep learning research in information retrieval.

Real-world scenarios involve tens of thousands (or even hundreds of thousands) of training queries with at least one positive label. Methods based on deep learning often require extensive training data, which has been a limitation for common information retrieval tasks like document ranking.

### 2.6.3 Natural Questions (NQ)

(Kwiatkowski et al., n.d.) Contains real anonymized queries issued to Google and annotated with relevant Wikipedia passages, suitable for passage ranking. It is designed to mimic the type of questions people naturally ask search engines and the type of answers they expect in return. The public release of the dataset consists of 307,373 training examples with single annotations, 7,830 examples with 5-way annotations for development data, and a further 7,842 examples 5-way annotated sequestered as test data. The NQ dataset is used for training models to answer both short and long form questions. In the paper (Kwiatkowski et al., n.d.) the authors demonstrate a human upper bound of 87% F1 on the long answer selection task, and 76% on the short answer selection task. Each example in the original NQ format contains the rendered HTML of an entire Wikipedia page, as well as a tokenized representation of the text on the page. A simplified version of the training set is also provided. The dataset is publicly available and can be accessed from the official Google Research Datasets GitHub page. It is also available on the Hugging Face Datasets library.

## 2.7 Loss Functions

In this section of the research study, we will discuss about various loss functions studied during literature survey. We discuss all the popular loss functions used in the domain of text ranking, text classification and text understanding. Going forward we will implement one of these loss functions for our training based on our dataset and model implementation.

### 2.7.1 Cross Entropy Loss

Cross entropy is a commonly used loss function for classification tasks. It measures the dissimilarity between predicted probabilities and ground truth labels. Specifically, for each pair of positive and negative examples, the CE loss encourages the model to assign higher probabilities to the positive example's class label. In the context of cross-encoders, CE loss aims to minimize the discrepancy between predicted relevance scores and true relevance labels.

### 2.7.2 Hinge Loss

Hinge loss is commonly used in support vector machines (SVMs) and ranking tasks. It encourages a margin between positive and negative examples. For cross-encoders, hinge loss penalizes the model when the predicted relevance score for a positive example is lower than that for a negative example. The hinge loss function is convex and promotes better separation between relevant and irrelevant pairs. (MacAvaney et al., 2019a) used hinge loss, also known as max margin loss, to fine-tune cross-encoders. This hinge loss is more commonly used when training neural re-rankers before the widespread adoption of pre-trained models like BERT, as evidenced in references (Hui et al., 2017)

### 2.7.3 Triplet Loss

Triplet loss is a technique used to train models to distinguish between similar and dissimilar things. It works by creating sets of three data points: an "anchor" point, a "positive" point that's like the anchor, and a "negative" point that's different from the anchor. The model is then trained to minimize the distance between the anchor and the positive point, while maximizing the distance between the anchor and the negative point. This approach is particularly useful for training Siamese networks (Koch et al., 2015) and FaceNet (Schroff et al., 2015), as well as contrastive learning techniques used in self-supervised learning.

### 2.7.4 Noise Contrastive Estimation (NCE) Loss

NCE loss transforms the problem of estimating a probability distribution into a binary classification task. The neural network is trained to distinguish between real data and a predefined fixed noise distribution. By doing so, NCE enhances efficiency and scalability, especially in high-dimensional and sparse data scenarios. The NCE loss introduced by (Gutmann and Hyvärinen, 2010) computes scores for both the positive instance and multiple negative instances. It then normalizes these scores into probabilities using the SoftMax function, with the goal of encouraging the model to assign a higher score to the positive instance compared to the negatives.

### 2.7.5 Localized Contrastive Estimation (LCE) Loss:

The LCE loss, proposed by (Gao et al., 2021) augments the NCE loss (Gutmann and Hyvärinen, 2010) with localized hard negative examples from the first-stage retriever. It combines the benefits of both contrastive estimation and hard negative mining. By selecting hard negatives from the retriever's output, LCE encourages the cross-encoder to distinguish between relevant and challenging-to-rank examples. (Pradeep et al., 2022) describes that LCE loss outperformed CE and hinge loss on the MS MARCO (Bajaj et al., 2016) passage ranking task, demonstrating its effectiveness for cross-encoders.

## 2.8 Summary

This chapter discussed the research conducted on various advanced encoder models and their training approaches, some of the researchers detailed the generation of synthetic training data. Major part of the research reviewed here is on negative sampling methods. The downstream task like retrieval/ranking of documents and passages was the primary focus of this literature review.

# Chapter 3 RESEARCH METHODOLOGY

## 3.1 Introduction

This section provides a detailed description of the methodology used in this study of augmenting existing data with hard negative mining and enhancing document ranking task. We will start with the proposed flow diagram, then dataset description and pre-processing to understand the characteristics of the data. This will be followed by data augmentation section. We will formulate the model architecture proposed in the methodology for hard negative mining and document re-ranking task as well as the evaluation metric going to be used after to training the model.

## 3.2 Proposed Flow

The proposed research methodology includes following steps:

- Selecting and loading required dataset.
- Exploratory data analysis of the dataset in hand.
- Pre-processing of dataset.
- Creating embedding representation of all the documents in the corpus.
- Perform clustering on top k documents related to the query.
- Select hard negatives from clustering result.
- Fine-tune ranking model with hard negatives.
- Evaluate and conclude the study.

The experiment pipeline is described in Figure3.1.

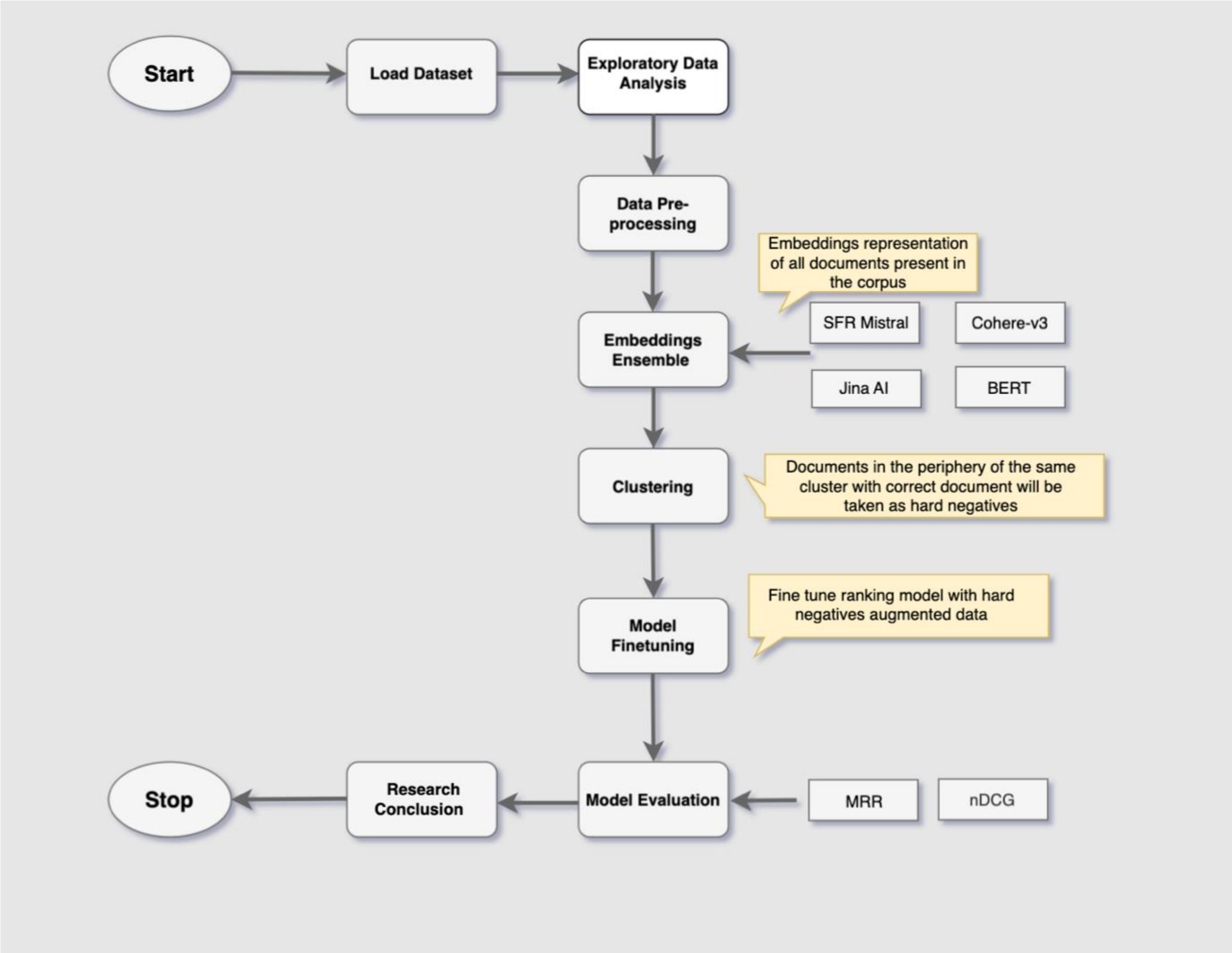

**Figure 3.1** Proposed flow of research methodology.

### 3.3 Dataset description and Pre-processing

In this research, the main aim is to enhance the re-ranking model performance on domain specific enterprise dataset. The research would capture and focus on the results on the private datasets of a company. The datasets will be split into train and test sets. For this research, we'll only use the test data to evaluate our results, just like other methods that are evaluated on similar dataset.

#### 3.3.1 Private Enterprise Dataset

Company's domain specific data is used in this research study. This data consists of URLs and full document, and it is well suited for document ranking task just like MS-MARCO (Bajaj et al., 2016) document ranking dataset. The documents are related to cloud domain and provide information about various cloud services and products. Enterprise has lots of documentation published in public or private web pages. The corpus of dataset contains 36,871 URLs and with pre-processing we must scrape the data from these URLs. The given corpus comprises documents of 30+ services/products. Figure 3.2 shows the distribution of the documents across services.

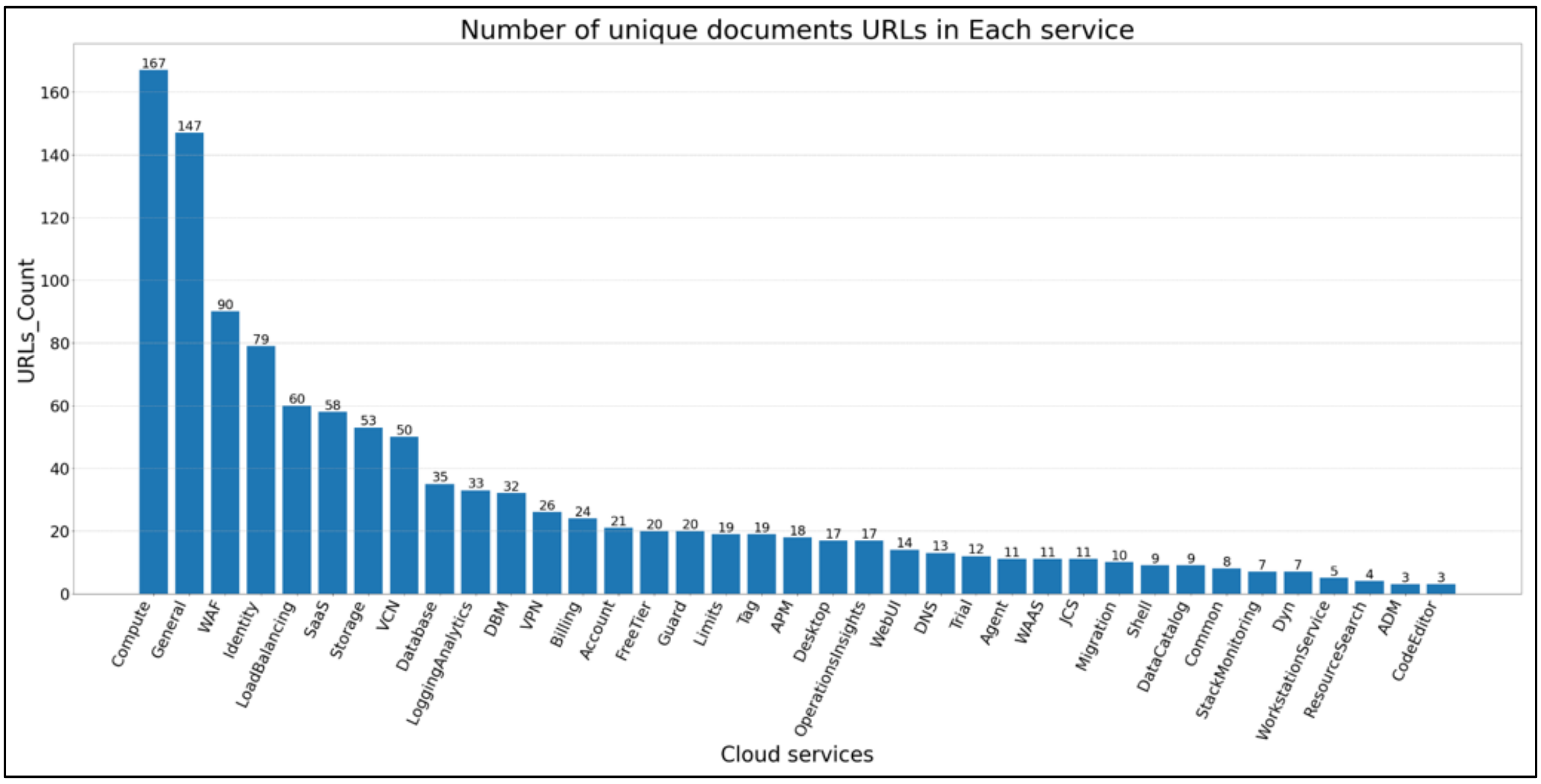

**Figure 3.2** Document distribution among different services and products.

### 3.3.2 Data Pre-processing

In this research, we aim to use enterprise dataset using documentation URLs. We will begin with document text extraction, data transformation and all text will undergo basic normalization steps including cleaning, spell correction and tokenization. We also focus on creating triplets for training once we get <query, document> pair, it is augmented with hard negatives.

#### *3.3.2.1 Text extraction from URLs*

The dataset consists of URLs of published web pages and text extraction is a very critical pre-processing step. We will use various web scraping tools and libraries such as Beautiful Soup, Selenium, Scrapy etc. and find out which one works best for our case.

- **Beautiful Soup:** Beautiful is a very powerful Python library for web scraping, especially for parsing HTML and XML documents. One of its most notable advantages is its convenience. It is very easy to use and navigate. It empowers you to try various techniques.
- **Selenium:** Selenium is a versatile web scraping library that automates web browsers, enabling interaction with dynamic web content. It supports multiple programming languages, including Python, and can handle JavaScript-heavy websites that static scrapers might miss. Selenium is often used for tasks like form submission, navigation, and data extraction from complex web pages. Its capability to mimic real user behavior makes it a powerful tool for web scraping.
- **Scrapy:** Scrapy is a robust Python framework for web scraping and web crawling. It allows developers to extract data from websites and process it as needed. Scrapy's architecture is designed for flexibility and scalability, making it suitable for large-scale data extraction tasks. Its extensive features and active community support make it a popular choice among developers.

*3.3.2.2 Data Transformation*

In the data transformation step, we transform the html/markdown text scraped from the URLs into the plain text format. The plain text format makes much more sense to humans as well as model. We use html2text library for converting html or markdown format data to text. We apply some custom cleaning to remove web page sidebars, headers and footers.

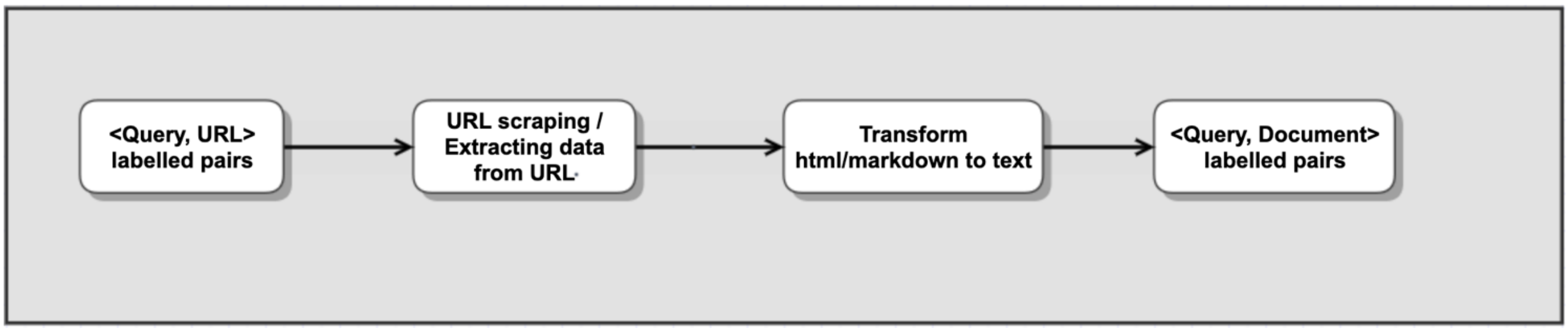

**Figure 3.3** Data transformation steps and flow diagram.

*3.3.2.3 Data augmentation with hard negative mining*

Augmenting data with hard negative is not just a pre-processing step but also a major part of the thesis to study and research how to find hard negatives which can boost model performance and help model to understand better the intricacies of domain specific dataset. We first generate embeddings for each document from different models, get the final one with ensemble and then perform clustering method to get the hard negative document for that query. A detailed explanation is provided in the next sections on hard negative mining.

## 3.4 Methodology

In this research, we will use multiple models to enhance our data with hard negatives. Initially, we will utilize multiple bi-encoder models to generate diverse embeddings for each document. These embeddings will then be combined using an ensemble technique to create a robust representation for each document. Following the embedding generation, clustering will be performed based on these embeddings. Using a similarity metric, we will identify additional positive documents for each query, as well as particularly challenging hard negative documents. This augmented dataset, enriched with hard negatives, will subsequently be used to train cross-encoder re-ranking models.

### 3.4.1 Embedding models (Bi encoders)

Bi-encoder models are pivotal in natural language processing (NLP) for creating document embeddings. These models independently encode input texts (e.g., queries and documents) into dense vector representations, or embeddings, in a shared latent space. This approach facilitates efficient similarity comparisons and retrieval tasks. A biencoder typically comprises two identical encoder networks, often based on transformer architectures such as BERT (Devlin et al., 2019a), RoBERTa (Liu et al., 2019). The encoders share weights, ensuring that the embeddings reside in a compatible semantic space, which is essential for measuring the relevance between the query and document vectors.

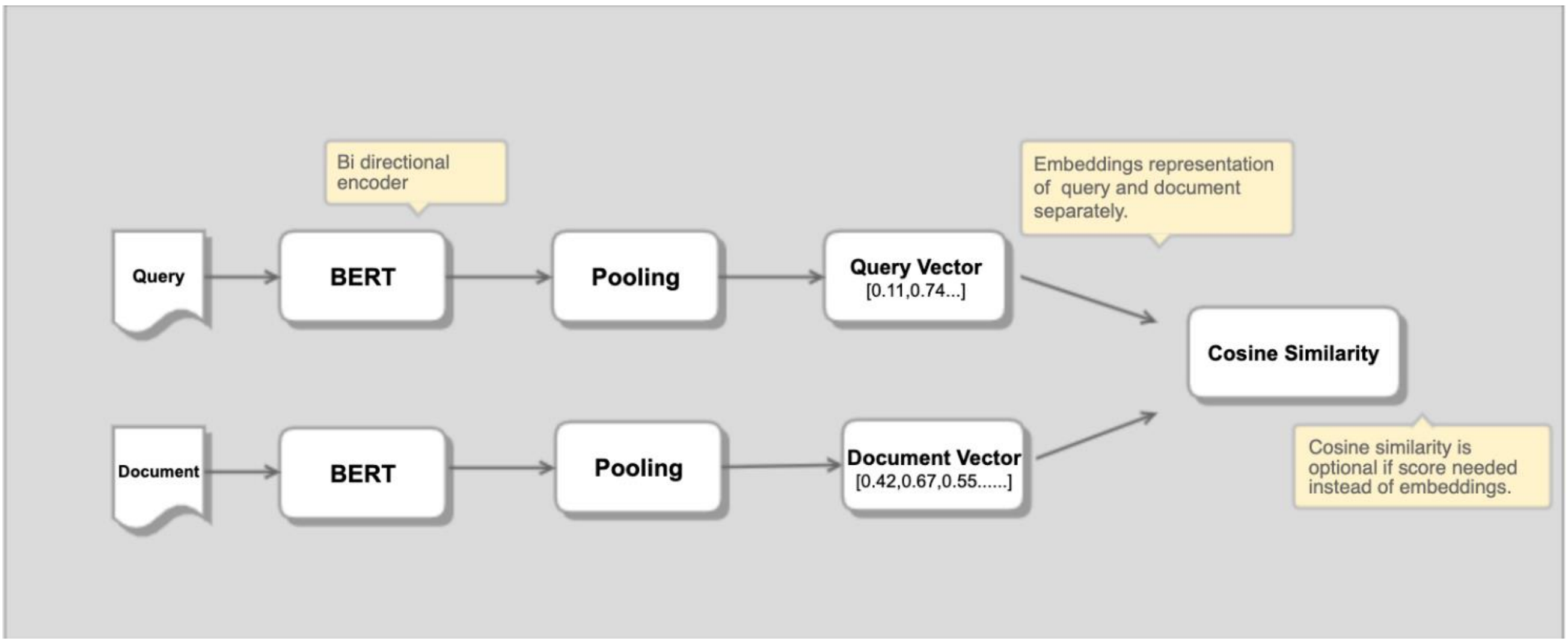

**Figure 3.4** Bi-encoder architecture

The primary advantage of biencoders lies in their computational efficiency during inference.
And hence, these are instrumental in scenarios requiring large-scale document retrieval, where the goal is to narrow down a vast pool of documents to a manageable subset before applying more computationally intensive models, such as cross-encoders, for fine-grained ranking. In the proposed methodology we propose to use pre trained state of the art embeddings models to create ensemble. Some of the embedding models are –

- Jina AI's embedding model: This is also available via Hugging Face transformer library. This embedding model supports sequence length of 8192, suitable for dataset which has long documents. The architecture is based on BERT (Devlin et al., 2019a)
- SFR Mistral Embedding: This model is by Salesforce Research. It is accessible through hugging face transformer library. It is trained on Mistral-7V-v0.1 and E5-mistral-7b-instruct.

- Cohere's embed-v3 model: We can use it via API endpoint, private deployment, AWS SageMaker or Hugging Face transformer library. The sequence length of this model is 512, either documents should be chunked within 512 tokens, or it will be truncated.

### 3.4.2 Ensemble Methods

Embeddings ensembles are a technique used to improve the performance of embedding models by combining multiple embedding models into a single model. This can be done in several ways, such as averaging the outputs of the different models, or using a more sophisticated voting scheme. There are several reasons why embeddings ensembles can be effective. First, different embedding models can capture different aspects of the data. By combining the outputs of multiple models, it is possible to get a more comprehensive representation of the data. Here's how we have applied voting ensemble to create robust embeddings similarity:

1. **Use Multiple Embedding Models:** Use different models to generate embeddings for all query and document dataset. These models can be of various types, such as:
   - Word embeddings (e.g., Word2Vec, GloVe)
   - Sentence embeddings (e.g., BERT, Sentence-BERT)
   - Custom neural network embeddings (e.g., jina_ai, gte-small)
2. **Generate Embeddings:** Use each model to generate embeddings for data points. This will result in multiple sets of embeddings for the same data points.
3. **Calculate Similarity Scores:** For each model, we calculate similarity scores between embeddings. Common similarity measures include cosine similarity, Euclidean distance, or dot product.
4. **Combine Similarity Scores using Voting:** We use a voting mechanism to combine the similarity scores from different models. There are two main types of voting we can use:
   - **Hard Voting:** Each model votes for a similarity result, and the majority vote is taken. In this case we can use hard voting for ranking of documents.
   - **Soft Voting:** The similarity scores are averaged or weighted to produce a final similarity score.

### 3.4.3 Clustering Algorithms

Clustering algorithms are essential for grouping similar document embeddings, enabling tasks such as topic modelling, document organization, and information retrieval. Here are some clustering algorithms particularly effective for handling document embeddings:

#### *3.4.3.1 K-Means Clustering*

K-Means is a widely used partitioning algorithm that clusters data into $k$k distinct groups based on feature similarity. Randomly initialize $k$ centroids. Assign each document embedding to the nearest centroid. Calculate new centroids as the mean of all embeddings assigned to each cluster. Repeat the assignment and update steps until centroids stabilize.

- Strengths
  - Simple and scalable.
  - Efficient for large datasets.
- Limitations
  - Requires predefining the number of clusters $k$.
  - Sensitive to initial centroid positions and outliers.

#### *3.4.3.2 Hierarchical Clustering*

Hierarchical clustering builds a tree of clusters either in a bottom-up (agglomerative) or top-down (divisive) manner. Treat each document embedding as a separate cluster. Iteratively merge the closest pair of clusters until a single cluster remains or the desired number of clusters is reached.

- Strengths
  - Does not require specifying the number of clusters in advance.
  - Produces a dendrogram for detailed cluster hierarchy.

- Limitations
  - Computationally intensive for large datasets.
  - Choosing the appropriate level to cut the dendrogram can be challenging.

#### *3.4.3.3 Gaussian Mixture Models (GMM)*

GMM assumes that document embeddings are generated from a mixture of several Gaussian distributions, using the Expectation-Maximization (EM) algorithm to estimate the parameters. Initialize the parameters (means, covariances, and mixing coefficients) of the Gaussian distributions. Use the EM algorithm to iteratively update the parameters and assign probabilities to each embedding for belonging to each Gaussian.

- Strengths
  - Provides a probabilistic clustering, offering flexibility in cluster assignment.
  - Can model clusters with different shapes and sizes.

- Limitations
    - Computationally intensive.
    - Requires specifying the number of clusters $k$

### 3.4.4 Re-ranking models (cross encoders)

Cross-encoder models are critical for re-ranking documents in domain-specific datasets. Unlike bi-encoder models, which independently generate embeddings for queries and documents, cross-encoders process the query and document together, allowing for a more diverse and context-aware evaluation of their relevance. In this approach, the query and document are concatenated and fed into the model as a single input sequence. This allows the cross-encoder to consider interactions between every token in the query and every token in the document simultaneously, capturing intricate dependencies and contextual variation that bi-encoders might miss.

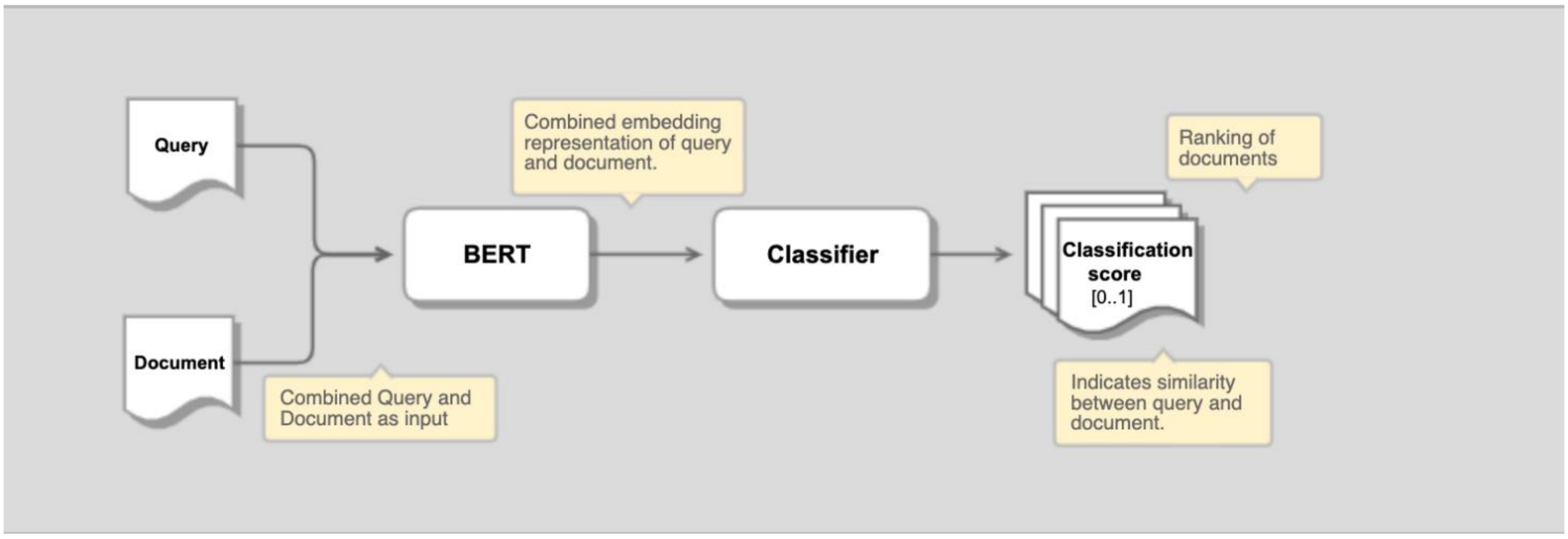

**Figure 3.5** Cross-encoder architecture.

However, cross-encoders are computationally intensive, as they require processing each query-document pair individually. This makes them less suitable for initial retrieval from very large datasets but ideal for applications where precision is paramount, such as legal document retrieval, medical literature search, or any specialized domain where understanding fine-grained details is critical. Some of the cross-encoder models are:

- distilroberta-base**:** DistilRoBERTa is a distilled version of RoBERTa (Liu et al., 2019) Distillation is a model compression technique where a smaller model (student) is trained to reproduce the behaviour of a larger model (teacher). DistilRoBERTa reduces the number of transformer layers by half compared to the original RoBERTa.

- Cohere rerank-english-v2.0: This model acts as a text relevance estimator. It takes a query and a collection of texts as input and outputs an ordered list, where each text is annotated with a relevance score calculated based on the provided query. The architecture is based on cross-encoder models (MacAvaney et al., 2019a)
- bge-reranker-base: bge stands for BAAI general embedding. Unlike embedding models, which produce intermediate vector representations, re-rankers directly compare questions and documents to output a relevance score. This score, however, can have any value due to the model's training using cross-entropy loss.

### 3.5 Evaluation Task and Metrics

In this research, the primary downstream task we aim to evaluate is document ranking. This involves assessing and ordering documents based on their relevance to a given query. We will explore various metrics commonly used in document ranking tasks to measure the effectiveness of our approach. These metrics may include precision, recall, F1 score, Mean Reciprocal Rank (MRR), and Normalized Discounted Cumulative Gain (NDCG), among others. Each metric provides a different perspective on the ranking performance, allowing us to comprehensively evaluate the quality and accuracy of the document ranking process.

#### 3.5.1 MRR

MRR is defined as the average of the reciprocal ranks of the first relevant document across multiple queries. The reciprocal rank is the multiplicative inverse of the rank at which the first relevant document appears.

Mathematically, MRR is given by:

$$MRR = \frac{1}{|Q|}\sum_{i=1}^{|Q|}\frac{1}{rank_i}$$

Where, |Q| is total number of queries and $rank_i$ is the rank position of the first relevant document for the $i^{th}$ query.

Steps to Calculate MRR -

1. Rank Documents: For each query, rank the documents in order of their relevance scores.
2. Identify Relevant Document: Determine the rank position of the first relevant document for each query.
3. Compute Reciprocal Rank: Calculate the reciprocal of the rank position for the first relevant document.

4. Average Reciprocal Ranks: Compute the mean of the reciprocal ranks across all queries to get the MRR.

MRR is a valuable metric for assessing the performance of document ranking tasks, providing insights into the effectiveness of ranking algorithms by focusing on the position of the first relevant document returned for each query.

### 3.5.2 Precision

Precision is a fundamental evaluation metric used in various information retrieval, machine learning, and classification tasks. It quantifies the accuracy of the positive predictions made by a model by measuring the proportion of true positive instances among all instances predicted as positive. Here we will use it to check how many are truly relevant documents in top K for example precision@3, precision @5 and precision@10.

$$Precision = \frac{TP}{TP + FP}$$

where, TP is actual document which is relevant to the query and FP are those documents which are fetched in top K instead of relevant documents.

## 3.6 Summary

This chapter thoroughly covers all the components required to implement the research approach. It includes the dataset, preprocessing procedures, algorithms to be used, and evaluation methods for assessing model performance.

# Chapter 4 IMPLEMENTATION AND ANALYSIS

## 4.1 Introduction

This section details the implementation of the approaches discussed in the research methodology. It covers data exploration, analysis, and experimentation, with the goal of discovering new insights and observations about the enterprise dataset. Detailed information about implementations and observations are explored in various subsections throughout this chapter.

## 4.2 Implementation

As shown in Figure 3.1, the proposed approach for research methodology includes pre-processing, data labelling, data augmentation, and modelling blocks, all integrated to form a comprehensive end-to-end pipeline. Additionally, the model architecture is detailed in Figures 3.4 and 3.5. This chapter will provide information about the implementation details of these models.

### 4.2.1 Data Pre-processing

This study focuses on augmenting enterprise dataset with challenging Hard Negative samples to train cross encoder re-ranker models for document ranking task. The data augmentation and pre-processing are distributed into four major tasks, data pre-processing for embedding models, ensemble approach for combining results of different embedding models, pre-processing for cross encoder models, and pre-processing for evaluating the document ranking task.

Before starting the data scraping and data transformation of given dataset, we will analyse the semantic words in context of the query. Let's see a visual representation of query where the size of each word indicates its frequency or importance in the dataset. It provides a quick and intuitive way to understand the most prominent words within a large corpus of text. Figure 4.1 indicates that the context of query consists of technical abbreviations like WAF, VCN, infrastructure. The domination of these words in the query makes it critical to utilize hard negative mining for models to learn specific context.

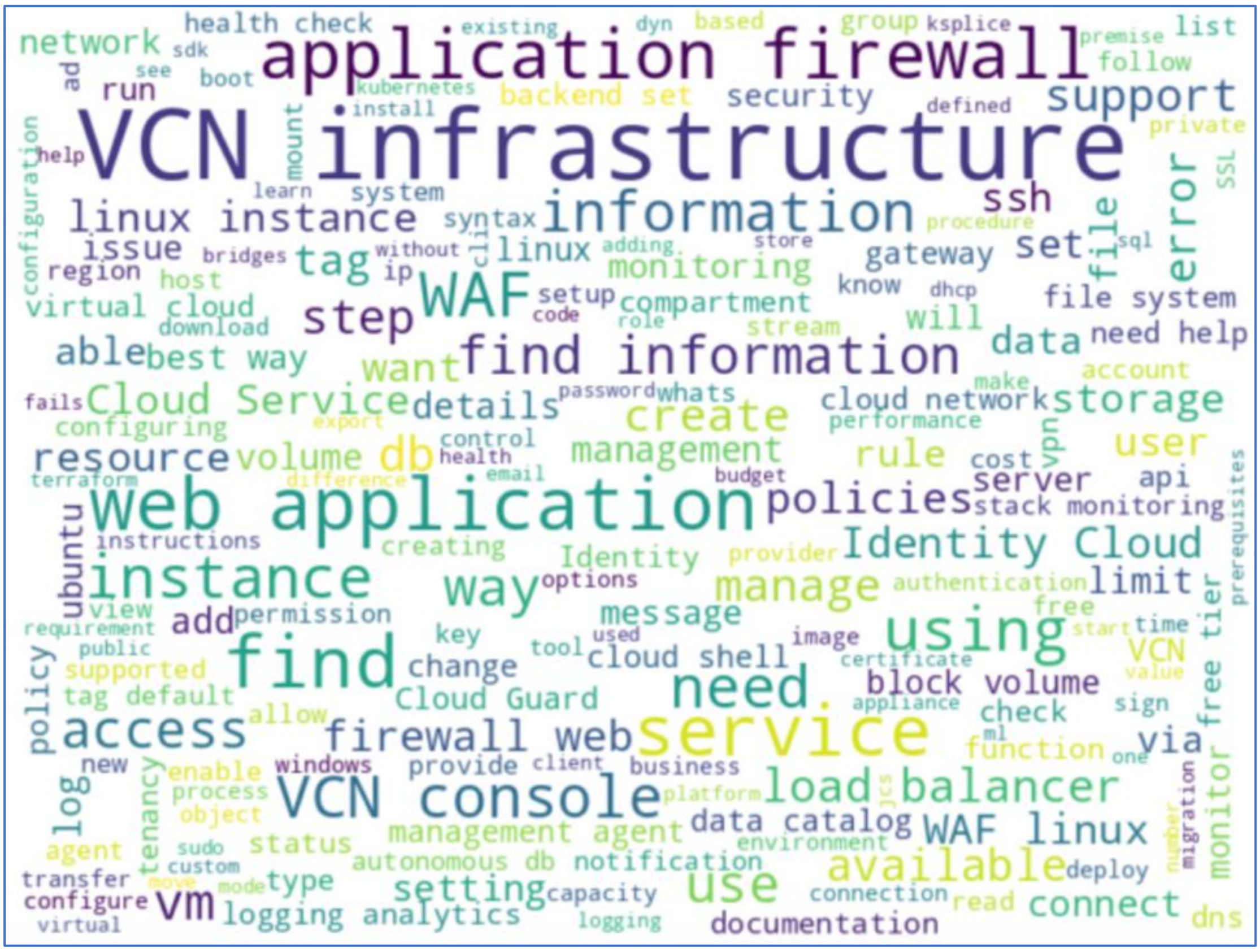

**Figure 4.1** Visual representation of common words in our enterprise dataset.

The implementation details of exploratory pre-processing to choose the useful information required for model training. Table 4.1 shows the examples of original dataset. The original dataset consists of the following columns:

- query_id: unique identifier for each query.
- query: these are frequently asked questions about a product or service created by subject matter experts.
- intent: this denotes the category of product or service associated with the query.
- service: this denotes the name of product or service associated with the query.
- url: This URL contains all the information about the corresponding query.
- doc_html: html format data extracted from the URL
- doc_text: After extracting data from URLs and converting it from html to text.

**Table 4.1** Example of original dataset

| query_id | query | intent | service | url |
|---|---|---|---|---|
| 426614174000 | How to upgrade my mount targets limit? | resourcediscoverymon.ans.RDMHostSupport | Limits | https://docs.cloud.com/iaas/Content/File/Tasks/creatingfilesystems.html |
| 426614173333 | How can I change the account administrator for my tenancy | resourcediscoverymon.ans.ChangeOcaToDomainController | Accounts | https://docs.cloud.com/en-us/iaas/process-automation/process-automation/create-iam-users.html |

Table 4.2 displays the data obtained after scraping URLs and converting the scraped data into text format, following the removal of unnecessary columns. The 'doc_html' column contains data extracted from URLs, while 'doc_text' consists of data converted into text. We use the URLs to scrape data employing the Scrapy library, and then transform the HTML data into text using html2text.

**Table 4.2** Example dataset after pre-processing

| query | url | doc_html | doc_text |
|---|---|---|---|
| How to upgrade my mount targets limit? | https://docs.cloud.com/iaas/Content/File/Tasks/creatingfilesystems.html | <div class= 'ResponseAnswer'> <p> Creating and Managing File systems and mount targets. Generally additional Mount Targets aren't needed since you can create 100 file systems per mount target. If security is an issue, NFS Exports can help with that. Please check out the following documentation:</p> target="_blank"href="https://docs.cloud.xyz.com/iaas/Content/File/Tasks/managingmounttargets.htm">Managing MountTargets</a></li><li><atarget="_blank"href="https://docs.cloud.xyz.com/iaas/Content/File/Tasks/exportoptions.htm">NFSExports</a></li></ul></div></div> | Creating and Managing File systems and mount targets. Generally additional Mount Targets aren't needed since you can create 100 file systems per mount target. If security is an issue, NFS Exports can help with that. Please check out the following documentation: Creating File Systems Managing Mount Targets NFS Exports - URL |
| How can I change the account | https://docs.cloud.com/en-us/iaas/process- | </div><div class='ResponseAnswer'><code class="language-plaintext">Please | Please Note: To change the admin, it is mandate that new admin user is added to |
| administrator for my tenancy | automation/process-automation/create-iam-users.html | Note: To change the admin, it is mandate that new admin user is added to the admin group of the tenancy and share us the screenshot the admin group or an email approval from the old admin/VP/Admins to proceed with the admin changes. </code></pre> <p> ................. How to create and add new user to admin group</a> guide and create a new local user account and add them to administrator group.</p> <p> For further help , please contact our chat agent with below details: </p> <li> Tenancy OCID/ Tenancy Name </li> <li> CSI No. or Subscription ID</li> <li>Admin email address</li> </ol> </div> </div> | the admin group of the tenancy and share us the screenshot the admin group or an email approval from the old admin/VP/Admins to proceed with the admin changes................................ How to create and add new user to admin group guide and create a new local user account and add them to administrator group. For further help , please contact our chat agent with below details: Tenancy OCID/Tenancy Name CSI No. or Subscription ID Admin email address |

From the above-mentioned features, we select “query” and “text_doc” for data augmentation using ensemble of embedding models. We will describe about ensemble and embeddings model in the next section. Based on our analysis for training a cross-encoder model, we augmented the data to select the necessary columns: “query,” “doc_text,” and “hard_negative_doc,”. Table 3.4 shows the number of columns after each step. With the required columns finalized, we then proceed to model-specific data pre-processing.

**Table 4.3** Dataset feature details after each pre-processing step

| **Original dataset columns** | **Columns after Pre-processing** | **Columns after data augmentation** | **Columns selected for model training** |
|---|---|---|---|
| query_id | query_id | query_id | query |
| query | query | query | doc_text |
| intent | intent | intent | hard_negative_doc |
| service | service | service | |
| url | url | url | |
| | doc_html | doc_html | |
| | doc_text | doc_text | |
| | | hard_negative_doc | |

### 4.2.2 Embedding Models for Data Augmentation

For selecting hard negatives, we utilize 2 different pre-trained embedding models. In this section, we will see the details of implementation of each of the embedding model. These models are taken from hugging face library.

- **jina-embeddings-v2-small-en:** The embedding model was initially trained with a sequence length of 512 but can extrapolate to lengths of 8k or more. This capability makes the model versatile for various use cases, particularly those involving long document processing, such as long document retrieval, semantic textual similarity, text reranking, recommendations, RAG, and LLM-based generative search. The model comprises 33 million parameters, enabling lightning-fast and memory-efficient inference while maintaining impressive performance. The jina-embeddings-v2-small-en is an English monolingual embedding model supporting an 8192-sequence length, based on the JinaBERT architecture.

```
=================================================================================
Layer (type:depth-idx)                                        Param #
=================================================================================
JinaBertModel                                                 --
├─JinaBertEmbeddings: 1-1                                     --
│    └─Embedding: 2-1                                         15,630,336
│    └─Embedding: 2-2                                         1,024
│    └─LayerNorm: 2-3                                         1,024
│    └─Dropout: 2-4                                           --
├─JinaBertEncoder: 1-2                                        --
│    └─ModuleList: 2-5                                        --
│    │    └─JinaBertLayer: 3-1                                4,198,912
│    │    └─JinaBertLayer: 3-2                                4,198,912
│    │    └─JinaBertLayer: 3-3                                4,198,912
│    │    └─JinaBertLayer: 3-4                                4,198,912
├─JinaBertPooler: 1-3                                         --
│    └─Linear: 2-6                                            262,656
│    └─Tanh: 2-7                                              --
=================================================================================
Total params: 32,690,688
Trainable params: 32,690,688
Non-trainable params: 0
=================================================================================
```

**Figure 4.2** Pre-trained embeddings model jina-embeddings-v2-small-en architecture

- **General Text Embeddings model (gte-small):** GTE models are built primarily on the BERT framework and are available in three sizes: GTE-large, GTE-base, and GTE-small. These models are trained on an extensive corpus of relevant text pairs across diverse domains and scenarios. This comprehensive training allows GTE models to be effectively utilized in various downstream tasks involving text embeddings, such as text reranking, information retrieval, semantic textual similarity, and more.

```
================================================================================
Layer (type:depth-idx)                                        Param #
================================================================================
BertModel                                                     --
├─BertEmbeddings: 1-1                                         --
│    └─Embedding: 2-1                                         11,720,448
│    └─Embedding: 2-2                                         196,608
│    └─Embedding: 2-3                                         768
│    └─LayerNorm: 2-4                                         768
│    └─Dropout: 2-5                                           --
├─BertEncoder: 1-2                                            --
│    └─ModuleList: 2-6                                        --
│    │    └─BertLayer: 3-1                                    1,774,464
│    │    └─BertLayer: 3-2                                    1,774,464
│    │    └─BertLayer: 3-3                                    1,774,464
│    │    └─BertLayer: 3-4                                    1,774,464
│    │    └─BertLayer: 3-5                                    1,774,464
│    │    └─BertLayer: 3-6                                    1,774,464
│    │    └─BertLayer: 3-7                                    1,774,464
│    │    └─BertLayer: 3-8                                    1,774,464
│    │    └─BertLayer: 3-9                                    1,774,464
│    │    └─BertLayer: 3-10                                   1,774,464
│    │    └─BertLayer: 3-11                                   1,774,464
│    │    └─BertLayer: 3-12                                   1,774,464
├─BertPooler: 1-3                                             --
│    └─Linear: 2-7                                            147,840
│    └─Tanh: 2-8                                              --
================================================================================
Total params: 33,360,000
Trainable params: 33,360,000
Non-trainable params: 0
================================================================================
```

**Figure 4.3** Pre-trained embedding model gte-small architecture

### 4.2.3 Ensemble Method and Clustering

The embeddings model will generate embeddings for each of the query and documents. For each model, we calculate similarity scores between embeddings. In this implementation we have use cosine similarity. For combining similarity score we use a voting mechanism on scores from different models. We use soft voting technique in which the similarity scores are averaged or weighted to produce a final similarity score.

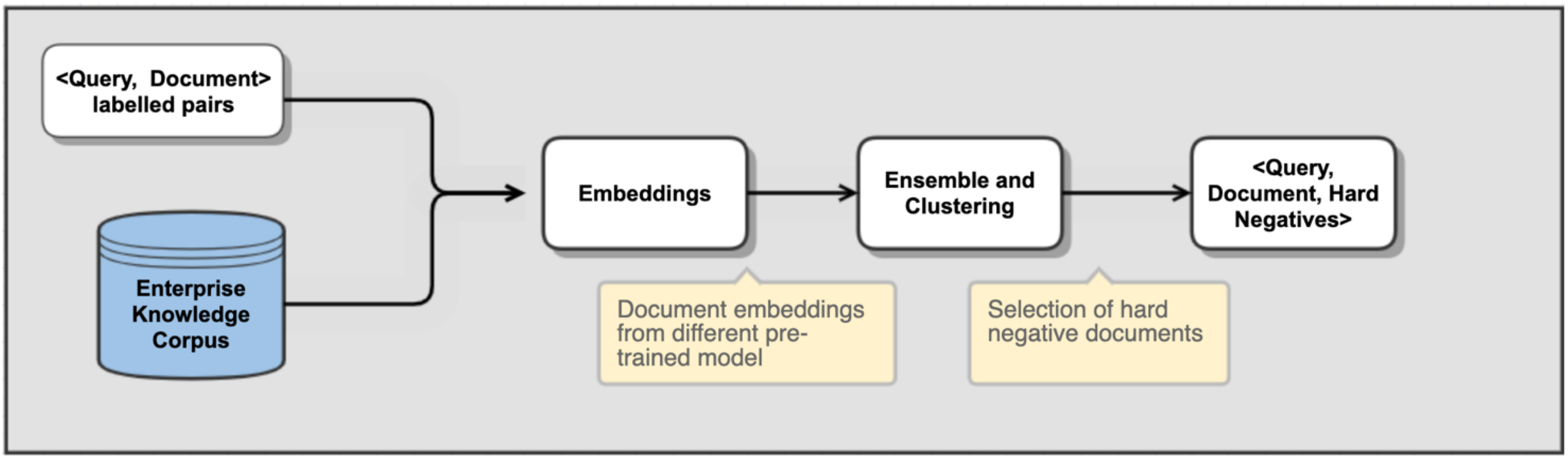

**Figure 4.4** Embeddings ensemble and clustering workflow for selecting hard negatives.

Figure 4.4 shows the hard negative mining process workflow. After getting scores from ensemble method, we use K-Means clustering technique on embeddings and similarity scores. Figure 4.5 shows how domain specific words are difficult to understand. Few things to observe in this plot –

- In the scatter plot, words that demonstrate a similarity exceeding a threshold of 0.5 are linked by dotted lines.
- In this context, VCN stands for Virtual Cloud Network, yet it shows less than 0.5 similarity. Conversely, VCN and VNIC display a similarity above 0.5, making VNIC a hard negative in this scenario.
- Similar is the case for WAF – Web Application Firewall. To mitigate this exact issue, we will train our cross-encoder on hard negatives specific to the domain of the dataset.
- This plot is a sample implementation of clustering with keywords instead of whole documents. In actual implementation, clustering is done on documents.

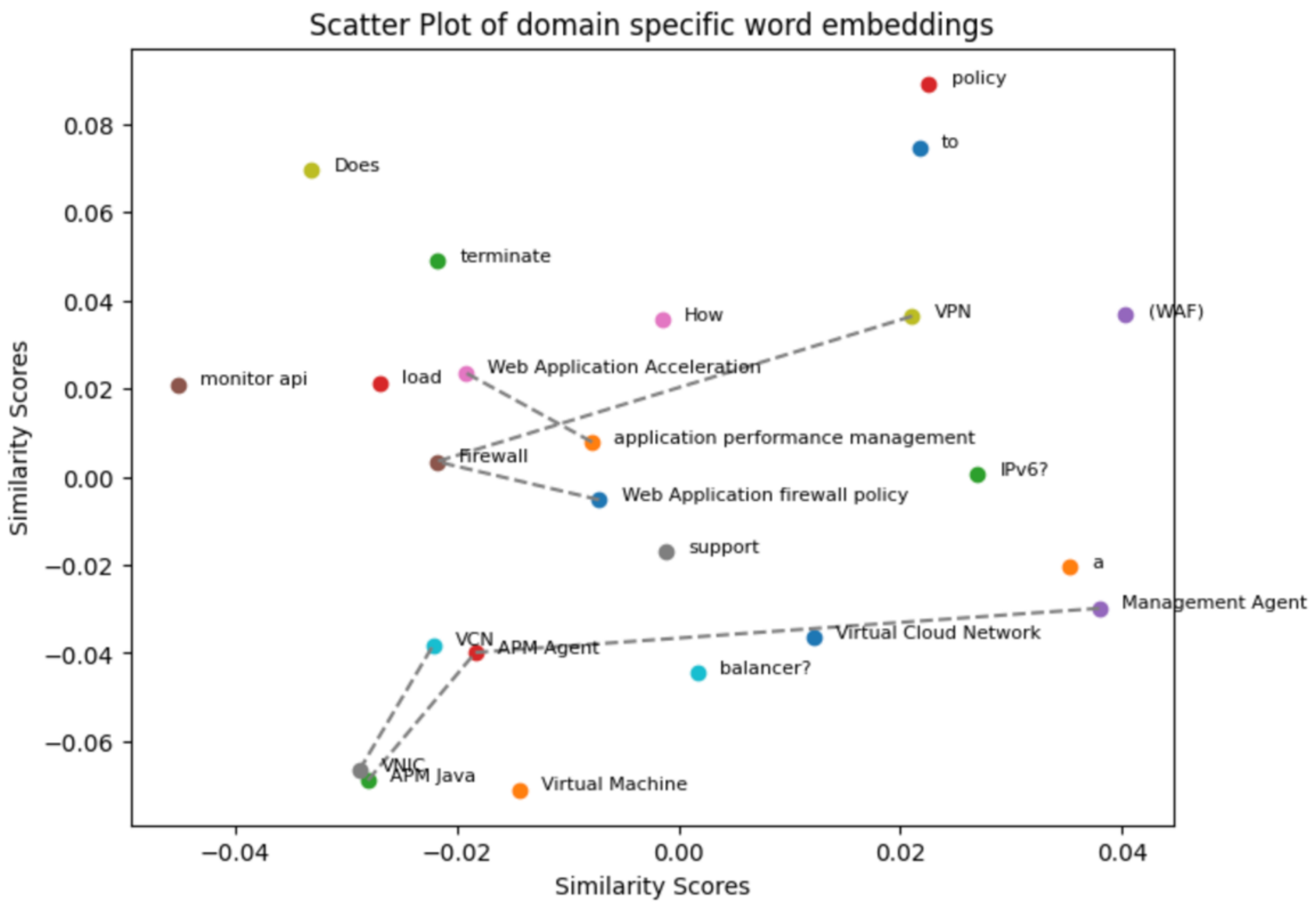

**Figure 4.5** Scatterplot based on similarity from single model.

### 4.2.4 Training of cross -encoder model

We utilize **distilroberta-base** model from the sentence transformer library as our base model, configuring it with num_labels = 1 to predict a continuous score ranging from 0 to 1. DistilRoBERTa is a distilled version of RoBERTa(Liu et al., 2019). Distillation is a model compression technique where a smaller model (student) is trained to reproduce the behavior of a larger model (teacher). DistilRoBERTa reduces the number of transformer layers by half compared to the original RoBERTa. It has 6 transformer layers instead of 12. The hidden size is maintained at 768. It uses 12 attention heads.

```
==========================================================================================
Layer (type:depth-idx)                                      Param #
==========================================================================================
RobertaForSequenceClassification                            --
├─RobertaModel: 1-1                                         --
│    └─RobertaEmbeddings: 2-1                               --
│    │    └─Embedding: 3-1                                  38,603,520
│    │    └─Embedding: 3-2                                  394,752
│    │    └─Embedding: 3-3                                  768
│    │    └─LayerNorm: 3-4                                  1,536
│    │    └─Dropout: 3-5                                    --
│    └─RobertaEncoder: 2-2                                  --
│    │    └─ModuleList: 3-6                                 42,527,232
├─RobertaClassificationHead: 1-2                            --
│    └─Linear: 2-3                                          590,592
│    └─Dropout: 2-4                                         --
│    └─Linear: 2-5                                          769
==========================================================================================
Total params: 82,119,169
Trainable params: 82,119,169
Non-trainable params: 0
==========================================================================================
```

**Figure 4.6** Cross-encoder model architecture trained for document re-ranking task.

**Table 4.4** Hyperparameters for cross- encoder reranker model

| **Parameter** | **Value** |
|---|---|
| Batch Size | 16 |
| Optimizer | Adam |
| Dropout | 0.2 |
| Learning rate | 3e-5 |
| Scheduler | Linear Scheduler with Warmup |
| Epochs | 20 |
| Loss | Triplet Loss |

The next steps involve transforming the dataset into a compatible format for the Sentence Transformers library. The model isn't capable of processing raw string lists directly. Therefore, each sample needs to be converted into the InputExample class format specific to the Sentence Transformers, and subsequently, the torch DataLoader class is utilized to batch and shuffle these examples. Specifically, the query, serving as the anchor, consists of a single sentence, while the pos (positive) contains a list of sentences, and neg (negative) comprises a list of selected hard negative sentences.

## 4.3 Analysis

We will analyse few exploratory aspects of data like distribution of documents across each service, document length distribution and query length distribution.

### 4.3.1 Queries Length Distribution

In this section we analyse the distribution of queries length in our enterprise dataset. Figure 4.7 shows that the length of queries ranges from 1 to 30 words, with some queries having very few words. This indicates that many documents could achieve high similarity scores with the queries due to the short query lengths and potentially longer document lengths. Therefore, when selecting hard negatives, it is crucial to consider not only the relationship between the query and documents but also the relationship between the positive document and other documents, ensuring a comparison with texts of nearly similar lengths. This finding can be considered while clustering the top similar documents with respect to query.

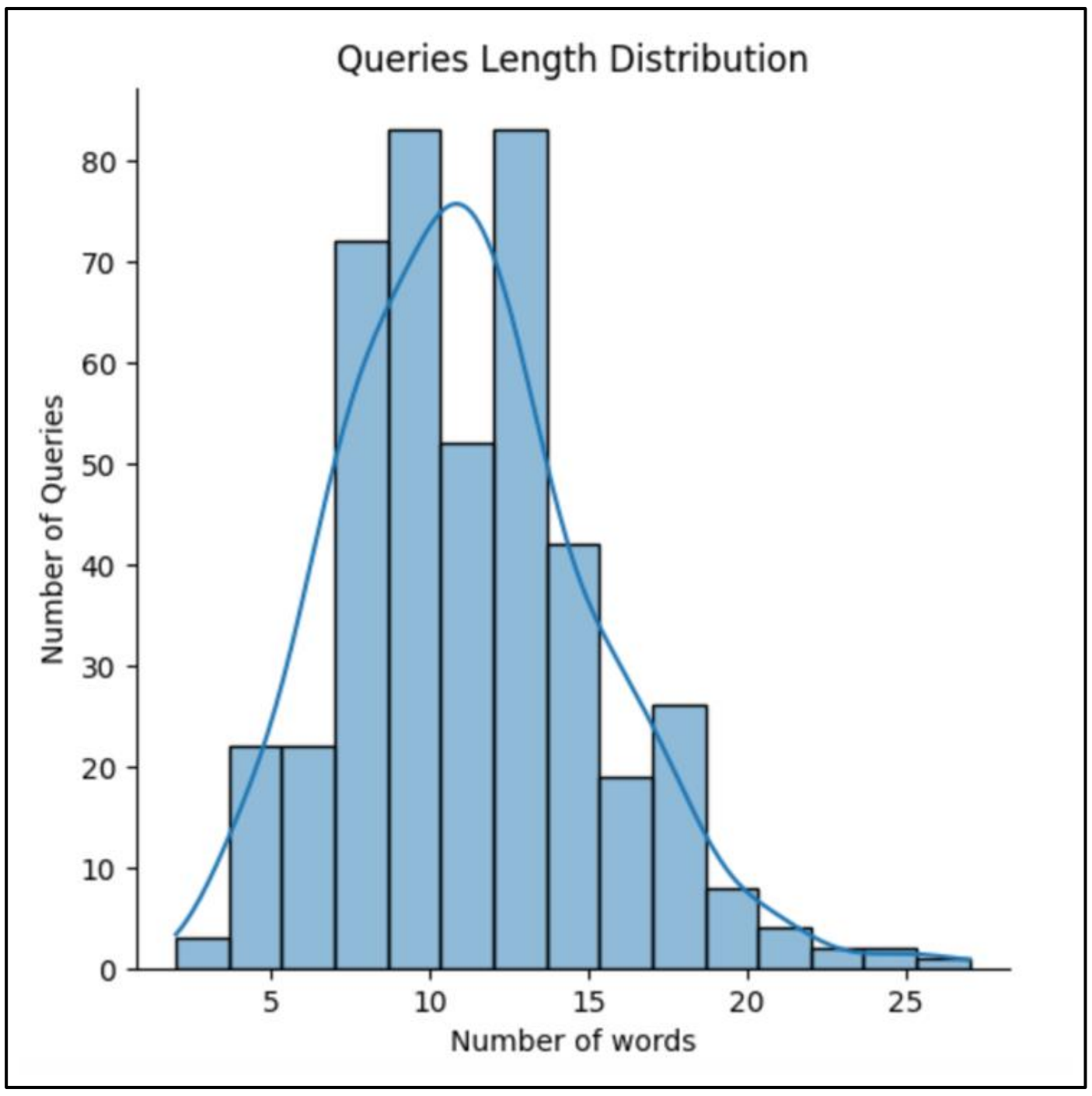

**Figure 4.7** Queries Length Distribution

### 4.3.2 Document Length Distribution

As shown in Figure 4.8, document lengths are significantly longer than query lengths. This disparity in context length affects the similarity scores, potentially reducing the accuracy of retrieval systems.

In the given dataset, each query is paired with a single correct document. This positive document is crucial for identifying challenging hard negatives and hence helpful for cross-encoder model training.

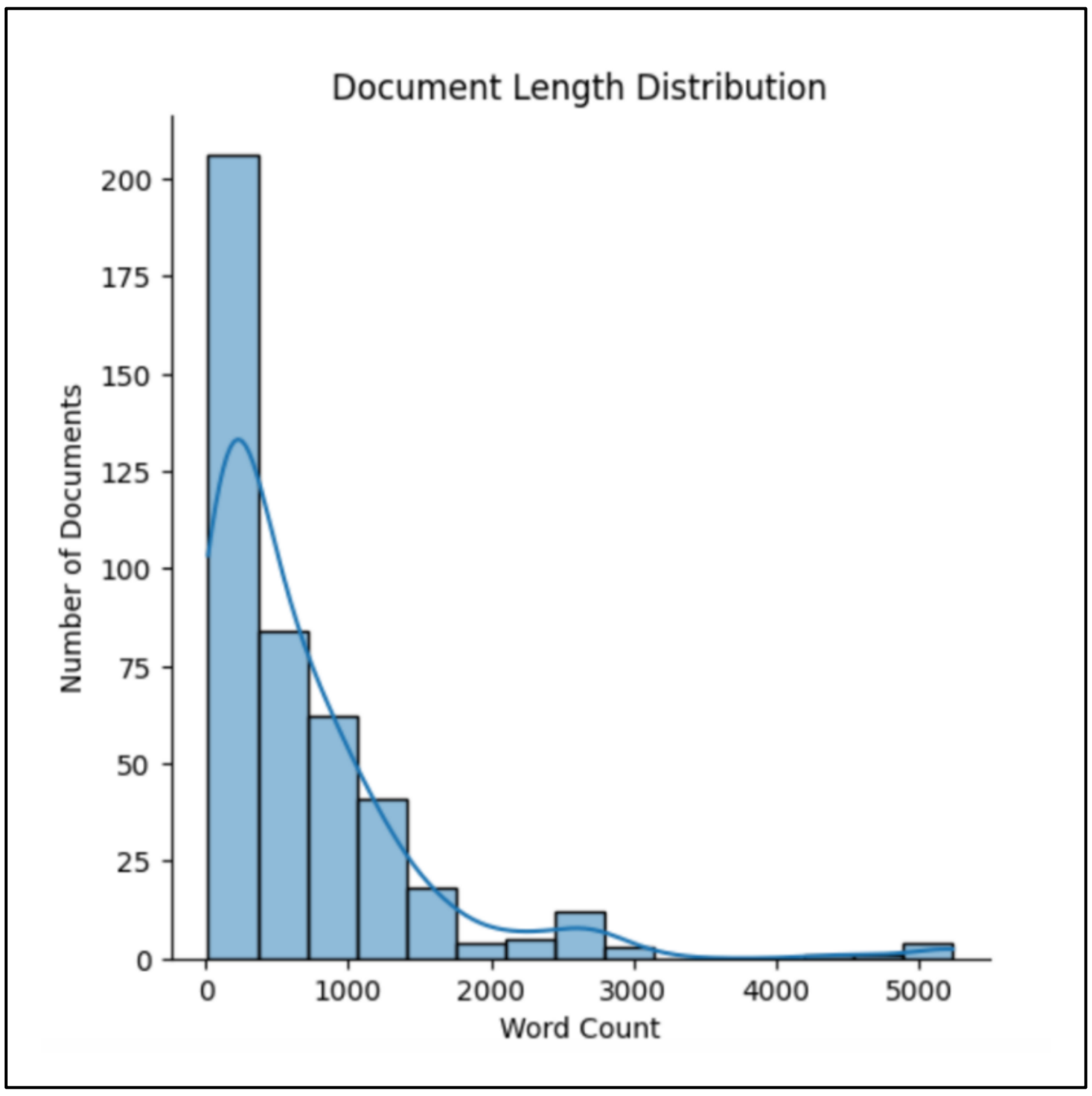

**Figure 4.8** Document Length Distribution

### 4.4 Framework requirements

In this research, we utilized several components from open-source projects, using them as a foundation for our own contributions to carry out various experiments and determine the final components. Table 4.3 outlines all the libraries employed in conducting these experiments.

**Table 4.5** List of open-source models and libraries used in this research experiment

| Open-source libraries | Description |
|---|---|
| Sentence Transformer | Cross-encoder reranker model. |
| Hugging Face | Pre-trained embedding models |
| PyTorch / PyTorch-Lighting | Model training and evaluation |
| Seaborn / Matplotlib | Visualization |
| Sckit-Learn | Helper function implementations |
| Tensorboard | For monitoring checkpoints and training loss |

This study focuses on selecting hard negatives using embeddings ensemble and ultimately training reranking model for enterprise dataset. The methodology involves large language models which requires heavy resource, and it is not possible to do in local infrastructure available. Hence for using GPU cloud compute resources were used. Table 4.4 lists the set of resources used.

**Table 4.6** List of hardware used in this research experiment.

| **Hardware** | **Description of Use** |
|---|---|
| Google Colab Pro | Platform and GPU to execute the experiments. |
| 1 A10 GPU | Embeddings representation, Model training |
| 150 GB Space | Temporary Disk space for Storing objects while the Instance is online |
| 100 GB Google Drive Space | Model checkpoints, Persistence saving Data |

## 4.5 Summary

In this chapter, we provided an in-depth discussion of the implementation details for the proposed pipeline components. We outlined the model architecture and listed the hyperparameters used. Additionally, we analyzed sequence length for representing queries and documents to understand embedding similarity metric. Finally, we also shared the framework required to continue experiments.

# Chapter 5 RESULTS AND DISCUSSIONS

## 5.1 Introduction

In this chapter, we will discuss the observed results for the task of document ranking on an enterprise dataset used in retrieval-based industry applications like Retrieval Augmented Generation (RAG). Section 5.2 will cover the results of using embedding ensembles and clustering techniques to select hard negatives. In Section 5.3, we will examine the outcomes of training the cross-encoder based re-ranker model on the enterprise dataset. Additionally, we will compare the results of training cross-encoders with random negatives against the proposed hard negatives. Finally, we will analyse the evaluation metrics and derive conclusions from the summary of all experiments.

## 5.2 Embeddings Ensemble and Clustering for Hard Negative Mining

Based on the Query Embedding, top 100 similar documents are selected from the vector store. We use both the embedding models separately to get 100 top similar ones. We merge all the retrieved documents to create a unique set of documents {Q1 : [D1,D2,D3….], Q2 : [D5,D7…] }

**Ensemble Scoring**: From the <Query Document> set create above, we create a matrix to score all the documents with each embedding model based on similarity/relevance. It creates a similarity matrix as shown in Table 5.1 for each of query. We then combine the scores by taking average of both the embedding models.

**Table 5.1** Shows the similarity matrix for every <Query, Document> pair based on each of the shortlisted embeddings.

| | **jina-embeddings-v2-small-en** | **gte-small** | **average score** |
|---|---|---|---|
| Q1D1 | 0.98 | 0.95 | 0.965 |
| Q1D2 | 0.87 | 0.67 | 0.77 |
| D1D3 | 0.67 | 0.77 | 0.72 |
| Q1D4 | 0.83 | 0.90 | 0.865 |
| Q1D5 | 0.84 | 0.75 | 0.795 |

After getting the combined scores from different embedding models, the next step is to cluster these document embeddings based on the ensembled similarity score. Before clustering, we need to tag the labelled positive document. The other remaining documents which have similarity score close to that of query and labelled positive, those documents are selected as hard negatives. Negative Documents are those having the least document score and Hard Negatives Documents are those having the highest similarity score with query and positive document both. Figure 5.1 illustrates the positioning of hard negatives within a cluster alongside the query and positive document. This visualization confirms that we have successfully identified hard negatives that were previously impairing the performance of ranking models due to their proximity to the query. If we neglect to train our model on these hard negatives, there is a significant risk that they may be ranked higher than the true positive documents, owing to unfamiliar jargons and technical implications.

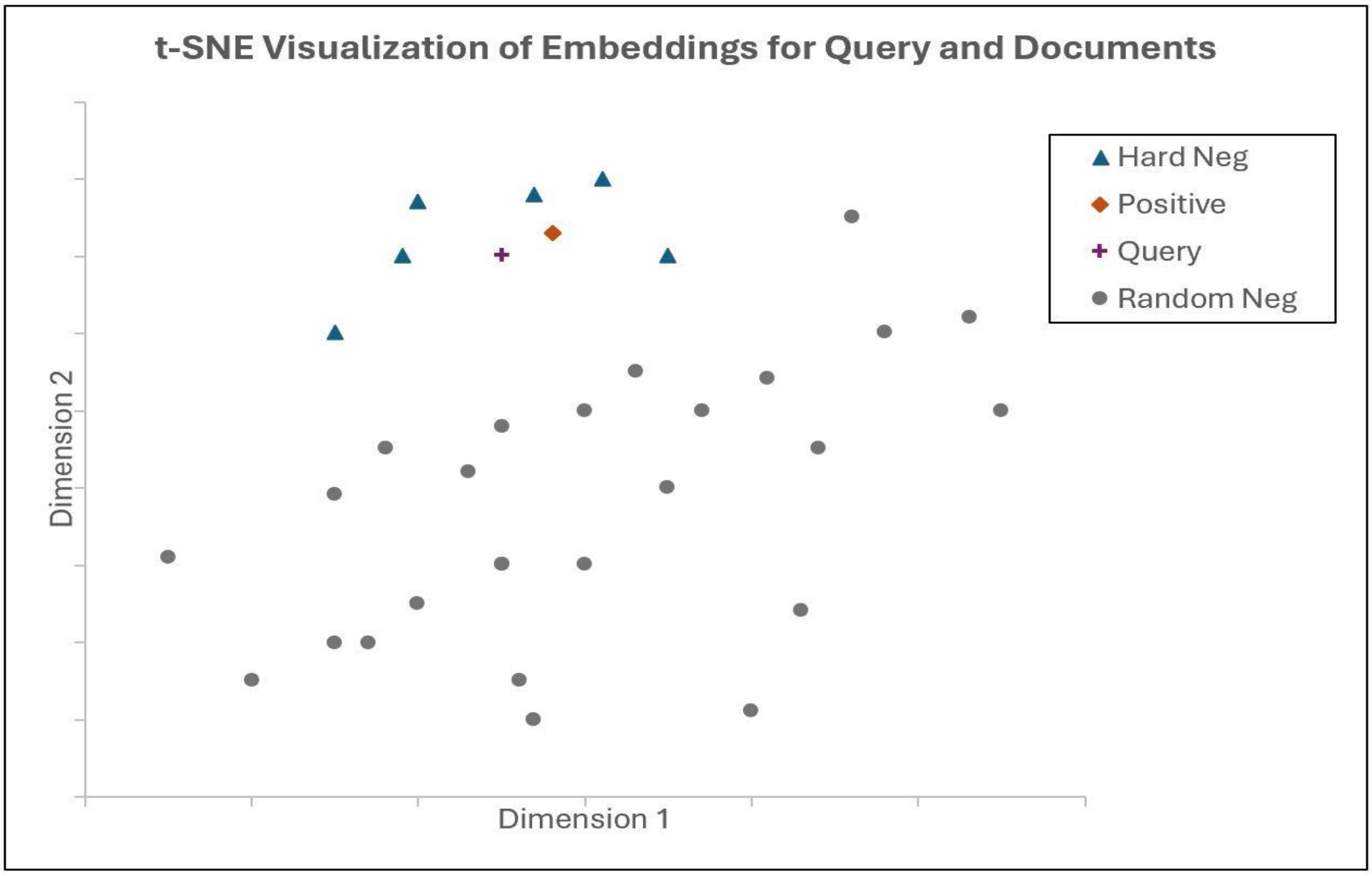

**Figure 5.1** Embeddings visualization of selected hard negatives in 2-dimension space with respect to query and positive document.

Our embeddings are high dimensional and t-SNE is better at creating a two-dimensional map that reveals the structure at many different scales. t-SNE is a non-linear dimensionality reduction technique. This is particularly useful for visualizing the data distribution of high-dimensional data. t-SNE is particularly effective at revealing clusters or groups within data.

## 5.3 Cross Encoder Re-ranker training Results

A Cross-Encoder takes a sentence pair as input and outputs a label. Here, it output a continuous label between 0 to 1 to indicate the similarity between the input pairs. In our experiment, we trained the distilroberta-base model on a triplet dataset that includes hard negatives, which were carefully selected using an embeddings ensemble. The trained Cross-Encoder model is particularly effective for tasks such as document re-ranking and passage re-ranking.

In practice, you can retrieve a set of 100 passages for a given query using a system like Elasticsearch. These retrieved passages, along with the query, are then fed into the Cross-Encoder, which scores them based on relevance. Finally, the passages are sorted according to the scores produced by the Cross-Encoder, ensuring the most relevant passages are ranked highest. This approach enhances the accuracy and effectiveness of information retrieval systems by leveraging the detailed semantic understanding provided by the Cross-Encoder model.

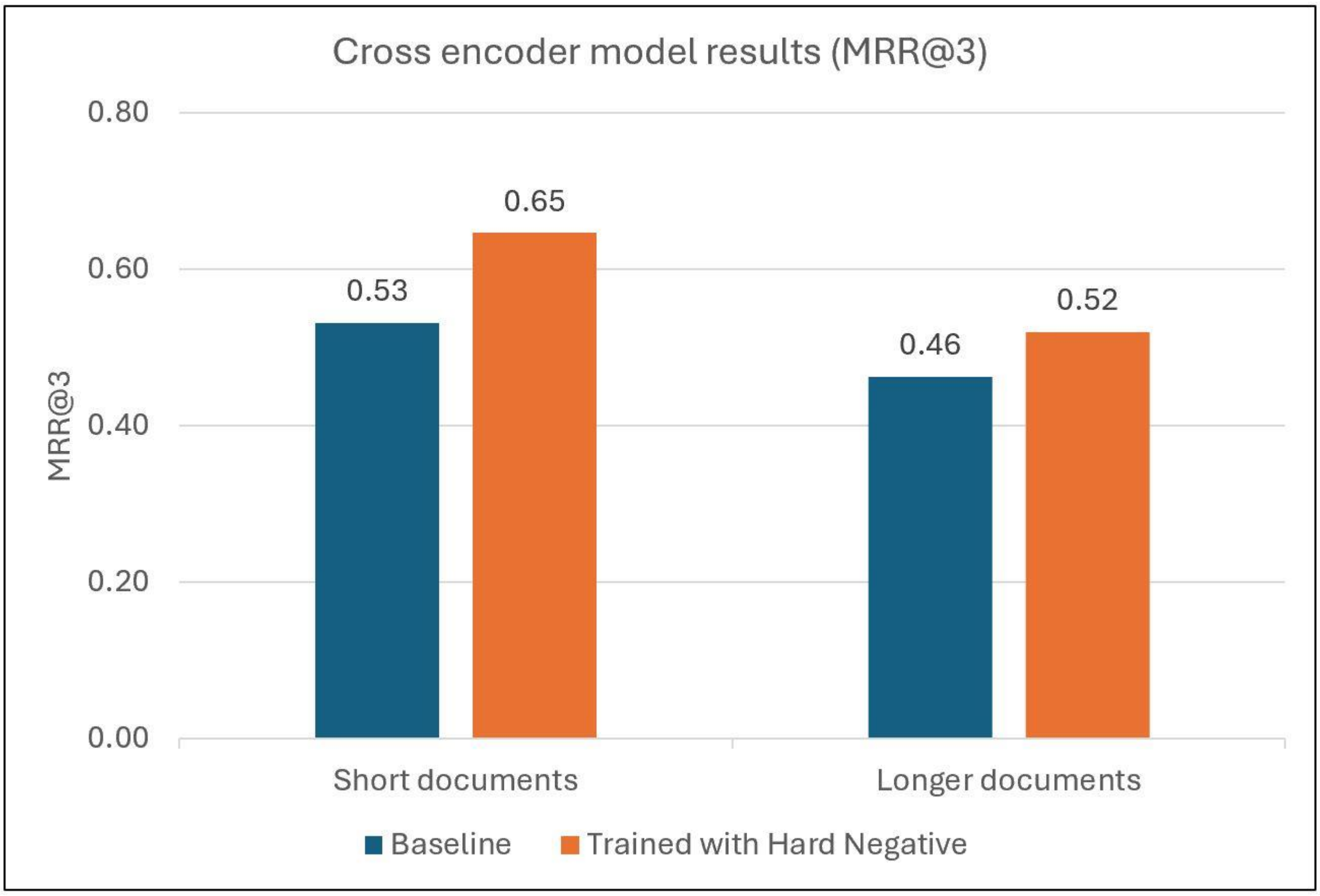

**Figure 5.2** MRR for the test split in two categories, short length documents (up to 2048 words) and longer documents (> 2048 words)

MRR measures the rank of the first relevant document. Higher MRR indicates better performance. We selected threshold of 2048 based on our analysis on document length distribution, majority of our documents lies in this range. From figure 5.2 we can observe and conclude that –

- Training with hard negatives significantly improved the MRR, as the model learns to distinguish between semantically similar documents more effectively.
- The reason for the improved performance of the fine-tuned model is that during training, each data point is presented with both the correct and incorrect options for the question. This allows the model to learn the semantic similarity between a query and its relevant answer, as well as the semantic dissimilarity between the query and hard negatives.
- This gives a significant boost compared to using random negatives in training data.
- Across both the baseline and fine-tuned models, as illustrated in Figure 5.1, there is a consistent trend where the MRR (Mean Reciprocal Rank) of long documents is significantly lower than that of short documents, with a disparity of approximately 9-10 points. This observation reinforces the issue that documents with high token counts are challenging to rank accurately against shorter queries.

Table 5.2 lists the resultant metric comparison of baseline and model trained on hard negatives of enterprise dataset. We have observed MRR@3, MRR@10, score average@3 and score average@10. MRR signifies the rank of the positive document. MRR@3 indicates where the positive document is ranked in top 3 documents, similarly we observe in top 10 documents. This order or ranking is based on the similarity score output of the cross-encoder. We also observe similarity score average of baseline and trained cross -encoder at top 3 and top 10.

**Table 5.2** Resultant metric comparison of baseline and proposed trained cross-encoder.

| Document length distribution | Model | MRR@3 | MRR@10 | sim_score average@3 | sim_score average@10 |
|---|---|---|---|---|---|
| Short documents | Baseline | 0.53 | 0.56 | 0.90 | 0.75 |
| | Trained with Hard Negatives | **0.64** | 0.64 | 0.61 | 0.40 |
| Long documents | Baseline | 0.46 | 0.51 | 0.86 | 0.71 |
| | Trained with Hard Negatives | **0.52** | 0.55 | 0.63 | 0.60 |

From Table 5.2 we can observe the following details:

For shorter documents:

- MRR@3 and MRR@10 show significant improvement.
- The average similarity score decreases because documents that were previously classified as positive with higher similarity scores might now fall into the hard negative category after training. As a result, these documents are no longer assigned higher scores, leading to a drop in the average similarity score.

For longer documents:

- Although there is some improvement in the MRR with training, the MRR remains low compared to shorter documents, even at the baseline.
- This may be due to the presence of extra information in long documents that may not be semantically like the query, thereby introducing noise that affects the overall document scores and rankings.
- Our embedding models have a limited context size, so tokens beyond the permitted size are truncated, potentially leading to the loss of important information. To address the challenges posed by longer documents, chunking strategies can be explored.

## 5.4 Limitations

This research highlights the use of different embedding models for selecting hard negatives for training data and fine-tuning the cross encoder on triplets of enterprise dataset to solve the task of document ranking, question answering and other retrieval tasks. Based on the experiments and implementation, we have identified some limitations of the study:

1. The enterprise dataset includes a corpus that allows for training embeddings from scratch. However, due to infrastructure limitations in our research study, we opted to use pretrained embedding models and applied an ensemble technique instead.
2. The embedding models utilized in this research are the smaller or base versions of state-of-the-art models. Using the full-scale versions of these models requires GPUs with substantial memory capacity. However, due to the memory limitations of the GPUs, we had to select and load embedding models that fit within the available memory capacity.
3. Many researchers generated synthetic data for hard negatives but since the dataset used here is very critical to enterprise customer experience, synthetic data generation was not explored.

4. Due to time constraints, chunking of long documents were out of scope. Performance of long documents can be further increased if proper chunking strategy is studies and implemented.

## 5.5 Summary

In this chapter, we have presented and discussed the results of using an embeddings ensemble to select hard negatives, as well as the outcomes of a fine-tuned cross-encoder on these proposed hard negatives. We showcased results for two categories of test data: short documents and long documents. It was observed that the fine-tuned cross-encoder on the triplet set outperforms the baseline cross-encoder. This improvement is due to the model's ability to learn both semantic similarity and dissimilarity simultaneously through Triplet Loss. Future work includes further enhancing the performance on long documents and exploring methods to train larger networks.

# Chapter 6 CONCLUSIONS AND RECOMMENDATIONS

## 6.1 Introduction

In this chapter, we summarize the research study by discussing the results and findings in alignment with the research aims and objectives. This study concludes and validates that hard negative mining can be effectively employed for document ranking tasks, particularly in the enterprise domain where domain-specific jargon may be unfamiliar to the initial embedding model. As NLP advances with large language model architectures built on vast datasets, this research provides a novel perspective to enhance retrieval models, ultimately influencing the performance of advanced LLM systems like Retrieval-Augmented Generation (RAG) and Reasoning and Action Agents (ReAct). Based on the experiments conducted, we propose fundamental recommendations for enhancements to support future research in this field.

## 6.2 Discussion And Conclusion

In this research study, the proposed approach uses hard negative mining by leveraging two distinct pre-trained embedding models to calculate document similarity scores. These scores are then combined to produce a robust and unbiased representation, especially since the documents contain unfamiliar enterprise-specific jargon. The ensembled similarity scores are subsequently clustered based on embedding similarity to identify hard negatives, considering their similarity with both the query and positive documents. Here, the pre-trained embedding models serve as the initial retrieval stage. The clustering results are used to select hard negatives for each query, and augmenting the existing data with these challenging hard negatives enables the model to match and even surpass baseline solvers suggested by researchers for the same task. This approach demonstrates that learning both similarity and dissimilarity simultaneously with cross-encoders improves performance. The findings indicate that a cross-encoder re-ranker trained using proposed hard negatives performs better than one trained with random or BM25 (Nguyen et al., n.d.) generated negatives.

The observed lower performance of large documents compared to short documents underscores the challenge of information lost in the middle and the inherent difficulty of processing such texts. Similar trends are evident in previously proposed baseline methods. This indicates that larger documents necessitate more advanced techniques for effective information retrieval. To enhance the performance on large documents, one potential approach is to segment the document into chunks based on semantic understanding or fixed sizes. Character-based chunking could also be beneficial.

This research demonstrates how document ranking can be enhanced using initial embedding models and hard negatives, focusing on learning <Query, Positive Document>, <Query, Negative Document> and <Positive Document, Negative Document> similarity and dissimilarity. The findings show that this approach yields competitive results compared to models proposed by other researchers. The experiments and evaluation metrics consistently indicate that the quality of hard negatives in triplets significantly improves the cross-encoder ranking performance on domain-specific documents containing unfamiliar terms.

## 6.3 Contribution to Knowledge

For Enterprises willing to adapt Generative AI for improving their search on internal documentations and downstream Question Answering tasks for customer support and chatbot use cases, this research is very beneficial. Through extensive experimentation, we demonstrate that even small adjustments and optimizations can lead to significant improvements in reranking effectiveness. The findings highlight that while cross-encoders are already powerful tools, there is still room for optimization that can yield substantial performance gains. This work emphasizes the importance of continual refinement and experimentation with existing models to push the boundaries of what they can achieve in practical applications.

This research enables industry practitioners and researchers to leverage data augmentation, embedding ensemble techniques, and hard negative mining on their industry-specific datasets for more effective semantic understanding, rather than training large language models (LLMs) for each downstream task. Although the dataset used in this study is an internal enterprise dataset, the code and understanding of this research will be open-sourced to facilitate further exploration and extension by the research community.

## 6.4 Future Recommendations

This research emphasizes the novel use of embedding ensembles for hard negative mining in training re-ranking models for enterprise search and document ranking tasks, an area that has not yet been extensively explored by the research community. Given the increasing interest in Retrieval Augmented Generation (RAG) applications across various industries, this approach offers significant potential and can be further developed. Based on our experiments, we suggest several future research directions that could build upon these findings.

1. To address the challenge of larger documents, one recommendation is to explore various chunking strategies during the document ingestion process.

2. Investigating why larger documents are more challenging to retrieve correctly compared to shorter documents and examining the impact on models and methods to bridge the performance gap between them.
3. We used soft voting ensemble method for embeddings scores, other ways of embedding's ensemble should be explored and benchmarked.
4. Additionally, cross-lingual and multilingual reranking systems could be explored to expand the applicability of these models globally.

# Appendix A: RESEARCH PLAN

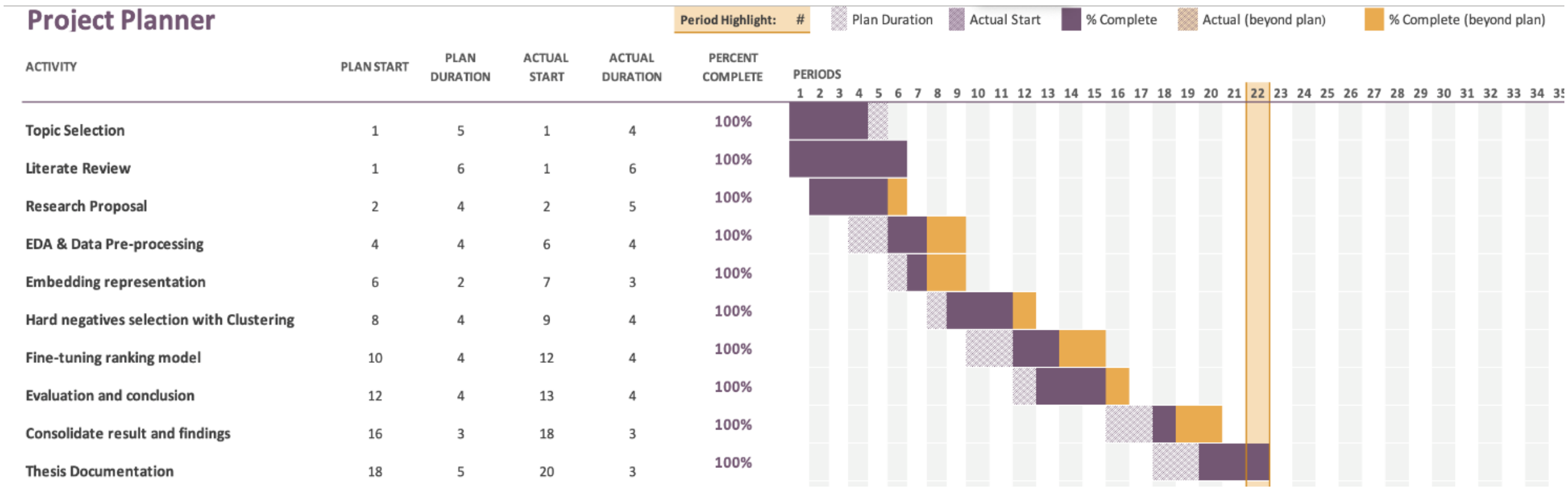

**Project Planner**

Period Highlight: # | Plan Duration | Actual Start | % Complete | Actual (beyond plan) | % Complete (beyond plan)

| ACTIVITY | PLAN START | PLAN DURATION | ACTUAL START | ACTUAL DURATION | PERCENT COMPLETE |
|---|---|---|---|---|---|
| Topic Selection | 1 | 5 | 1 | 4 | 100% |
| Literate Review | 1 | 6 | 1 | 6 | 100% |
| Research Proposal | 2 | 4 | 2 | 5 | 100% |
| EDA & Data Pre-processing | 4 | 4 | 6 | 4 | 100% |
| Embedding representation | 6 | 2 | 7 | 3 | 100% |
| Hard negatives selection with Clustering | 8 | 4 | 9 | 4 | 100% |
| Fine-tuning ranking model | 10 | 4 | 12 | 4 | 100% |
| Evaluation and conclusion | 12 | 4 | 13 | 4 | 100% |
| Consolidate result and findings | 16 | 3 | 18 | 3 | 100% |
| Thesis Documentation | 18 | 5 | 20 | 3 | 100% |

# APPENDIX B: RESEARCH PROPOSAL

## Abstract

Ranking consistently emerges as a primary focus in information retrieval research. Retrieval and ranking models serve as the foundation for numerous applications, including web search, open domain QA, enterprise domain QA, and text-based recommender systems. Typically, these models undergo training on triplets consisting of binary relevance assignments, comprising one positive and one negative passage. However, their utilization involves a context where a significantly more nuanced understanding of relevance is necessary, especially when re-ranking a large pool of potentially relevant passages. Although collecting positive examples through user feedback like impressions or clicks is straightforward, identifying suitable negative pairs from a vast pool of possibly millions or even billions of documents possess a greater challenge. Generating a substantial number of negative pairs is often necessary to maintain the high quality of the model. Several approaches have been suggested in literature to tackle the issue of selecting suitable negative pairs from an extensive corpus. This study focuses on explaining the crucial role of hard negatives in the training process of cross-encoder models, specifically aiming to explain the performance gains observed with hard negative sampling compared to random sampling. We would like to develop a novel hard negative sampling technique for efficient training of cross-encoder re-rank models on a custom data which has domain specific context.

## 2. Background

Retrieval models, unlike generative models, fetch real information from sources, with search engines indicating the source of each retrieved item (Sanderson and Croft, 2012). This underscores the continued importance of information retrieval (IR), even in the presence of generative LLMs, particularly in contexts where reliability is crucial. After the launch of BERT (Devlin et al., 2019b) in October 2018, a straightforward retrieve then re-rank strategy became popular in January 2019 as a successful means of leveraging pre-trained transformers for passage retrieval (Nogueira and Cho, 2019b). This model, known as monoBERT, marks the initial manifestation of what later evolved into cross-encoders for retrieval, a category encompassing reranking models such as MaxP (Dai and Callan, 2019), CEDR (MacAvaney et al., 2019b), Birch (Akkalyoncu Yilmaz et al., 2019), PARADE (Li et al., 2020), and numerous others. A proficient ranking algorithm holds potential advantages for numerous downstream tasks within information retrieval research (Han et al., 2020). Conventional algorithms like BM25 (Robertson and Walker, 1994) heavily rely on term-matching metrics, limiting their efficacy to scenarios where queries and documents share identical terms. This inherent drawback leads to performance degradation when faced with semantic differences despite identical meanings, a phenomenon known as the vocabulary mismatch problem. To enhance the understanding of users' search intentions and retrieve pertinent items, it's anticipated that ranking algorithms would engage in semantic matching between queries and documents (Li and Xu, 2014). Driven by advancements in deep learning, particularly techniques for capturing meaning (representation learning), researchers are increasingly using Dense Retrieval (DR) models to tackle the challenge of semantic similarity (Guu et al., 2020; Karpukhin et al., 2020; Luan et al., 2020). DR excels at capturing the semantic essence of queries and documents by converting them into low-dimensional embeddings. This facilitates efficient document indexing and similarity search, leading to effective online ranking. Studies have shown promising results for DR models in various information retrieval tasks. (Guu et al., 2020; Karpukhin et al., 2020; Qu et al., 2020)

While past research using diverse training strategies for DR models has shown encouraging results, inconsistencies and even contradictions arise when comparing their findings. For example, the superiority of training with "hard negatives" (highly similar yet irrelevant documents) over random negatives remains an open question. Additionally, many effective training methods suffer from inefficiency, making them impractical for large-scale deployments. Despite promising results, DR faces key challenges, and we are trying to investigate one of the challenges related to hard negatives mining on a custom data which has specific industry context.

## 3. Problem Statement and Related Research

The major issues while training the retrieval models -

- Hard negatives, which are non-relevant passages that closely resemble positive examples, play a crucial role in refining the model's understanding.
- Providing both positive (relevant) and negative (irrelevant) examples is important. Negative examples, especially hard negatives, challenge the model to distinguish between relevant and irrelevant content effectively.
- When dealing with enterprise-specific datasets, the inclusion of hard negatives is paramount.
- Without exposure to hard negatives, the model might struggle to differentiate between similar passages, leading to inaccurate responses and compromised decision-making processes within organizations.

### 3.1 Hard Negative

Previous studies highlight the critical role of negative example selection in optimizing the training process of cross-encoder models. (Karpukhin et al., 2020) conducted a study to assess the efficacy of different negative sampling strategies in training. They compared the performance of, BM25 (Robertson and Walker, 1994) negatives, random negatives, and in-batch negatives. Their findings suggest that combining BM25 and in-batch negatives leads to the most favorable outcomes in terms of training effectiveness. (Xiong et al., 2020) theoretically demonstrate that employing local negatives is suboptimal for dense retrieval learning. Subsequently, they suggest a method for generating global negatives by utilizing the existing dense retrieval model concurrently during training. This approach necessitates periodic re-indexing of the corpus and retrieval. (Qu et al., 2020) additionally suggest a method for generating hard negatives by utilizing the existing dense retrieval model, albeit after the completion of training rather than dynamically during training. However, their study reveals that employing these hard negatives alone post-training could potentially hinder the training process. They note that the effectiveness of these hard negatives is improved when filtered based on an independently trained cross-encoder model. (Zhan et al., 2021) discover that the instability induced by hard negatives can be mitigated by incorporating random negatives into the training process. Furthermore, they adopt a strategy akin to the ANCE (Xiong et al., 2020) method for periodically re-generating hard negatives, with the modification of updating only the query encoder to reduce re-indexing overhead. These findings underscore the significance of hard negatives in training and demonstrate varying levels of effectiveness across different approaches. Apart from the aforementioned research, which primarily delves into hard negative training strategies for fine-tuning DPR-like bi-encoders, other studies have made comparable observations regarding diverse methodologies that benefit bi-encoders. (Gao and Callan, 2021) highlight the continued importance of hard negatives even when the model undergoes additional pretraining aimed at enhancing the representation of the [CLS] token, referred to as Condenser in their study. The significance of hard negative mining has also been demonstrated in scenarios where knowledge distillation is implemented on bi-encoders. (Hofstätter et al., 2021; Lin et al., 2021) In contrast to the numerous studies investigating the benefits of hard negatives for bi-encoders, only one study (Gao et al., 2021) successfully integrates hard negatives into cross-encoder training. They introduce the Localized Contrastive Estimation (LCE) loss to illustrate its efficacy, demonstrating that integrating this loss function with harder negatives substantially enhances reranking performance, particularly when the distribution of training instances aligns with the results retrieved by initial-stage retrievers.(Pradeep et al., 2022)

### 3.2 Cross encoder

The initial cross-encoder for reranking, monoBERT (Nogueira et al., 2019), swiftly emerged shortly after the introduction of BERT (Devlin et al., 2019a) It adhered to the approach advocated by the BERT authors for processing (query, document) input pairs and exhibited substantial advancements in effectiveness on datasets such as TREC CAR (Dietz et al., n.d.) and MS MARCO passage ranking (Bajaj et al., 2016). While the vanilla version of monoBERT (Nogueira et al., 2019) demonstrated significant enhancements in the passage retrieval task, its design did not accommodate the processing of long input sequences necessary for document retrieval. Many subsequent BERT-based cross-encoder studies (MacAvaney et al., 2019a; Li et al., 2020) aimed to tackle this challenge by either conducting multiple inferences on various segments of the document or implementing additional architectural modifications atop BERT to enhance the processing of longer document texts.

## 4. Aim and Objectives

The principal aim of this research is to propose a hard negative mining strategy which helps enterprises to fine-tune state of the art ranking/retrieval models on their private dataset. These datasets are unique to each enterprise and may contain specialized terminologies and domain-specific jargon. So, by incorporating hard negatives in the training process, retrieval models can be honed to navigate the complexities of enterprise data, ensuring precise and contextually relevant results tailored to the specific needs of the business.

The research objectives are derived from the purpose of this study, which are outlined as follows:

- To suggest a robust hard negative mining strategy for domain specific private data.
- To utilize the hard negatives for fine-tuning of cross-encoder model on domain specific data.
- To investigate the impact of using hard negatives in training of ranking models.

## 5. Significance of the Study

Large Language Models (LLMs) excel at answering questions based on the knowledge they were trained on. However, their training data typically does not include recent information and specific private information stored in platforms like a company's Confluence, Google Drive, or SharePoint. For the integration of Generative AI within enterprise settings, every organization will require a custom retrieval system built upon their private datasets. Retrieval augmented systems, when not trained on enterprise domain datasets, face significant challenges in delivering accurate and relevant results within organizational contexts. One of the primary issues is a lack of familiarity with the specific terminology, jargon, and nuances prevalent in the enterprise domain. Importance of training retrieval/ ranking models with hard negatives cannot be overstated, especially within enterprise domains. This study bridges the gap between pre-trained ranking models and domain-specific data. The study also proposes a data augmentation technique to include hard negatives for given query, document pair. This study will investigate that the triplet objective function can be utilized by researchers across various NLP applications. Additionally, it aims to introduce a modified evaluation criterion suitable for use by researchers evaluating ranking models for internal dataset.

## 6. Scope of the Study

The scope of the study is to develop hard negative mining strategy using ensemble of embeddings. Pre-trained embedding models used here will be state of the art models like SFR-Embedding-Mistral, Jina AI, Cohere-v3 etc. And to analyse impact of fine-tuning cross encoder models with mined hard negatives on the public dataset MS MARCO as well as private enterprise dataset. Given resource limitations, this study will not explore training embedding models from scratch using large-scale text corpora. We will instead leverage pre-trained embeddings and adapt the size and complexity of candidate models to fit within available resources. Training models approaching the scale of leading-edge architectures is not feasible for this study.

## 7. Research Methodology

### 7.1 Introduction

The proposed research methodology includes following steps:

- Selecting and loading required dataset.
- Exploratory data analysis of the dataset in hand.
- Pre-processing of dataset.
- Creating embedding representation of all the documents in the corpus.
- Perform clustering on top k documents related to the query.
- Select hard negatives from clustering result.
- Fine-tune ranking model with hard negatives.
- Evaluate and conclude the study.

The experiment pipeline is described in Figure1.

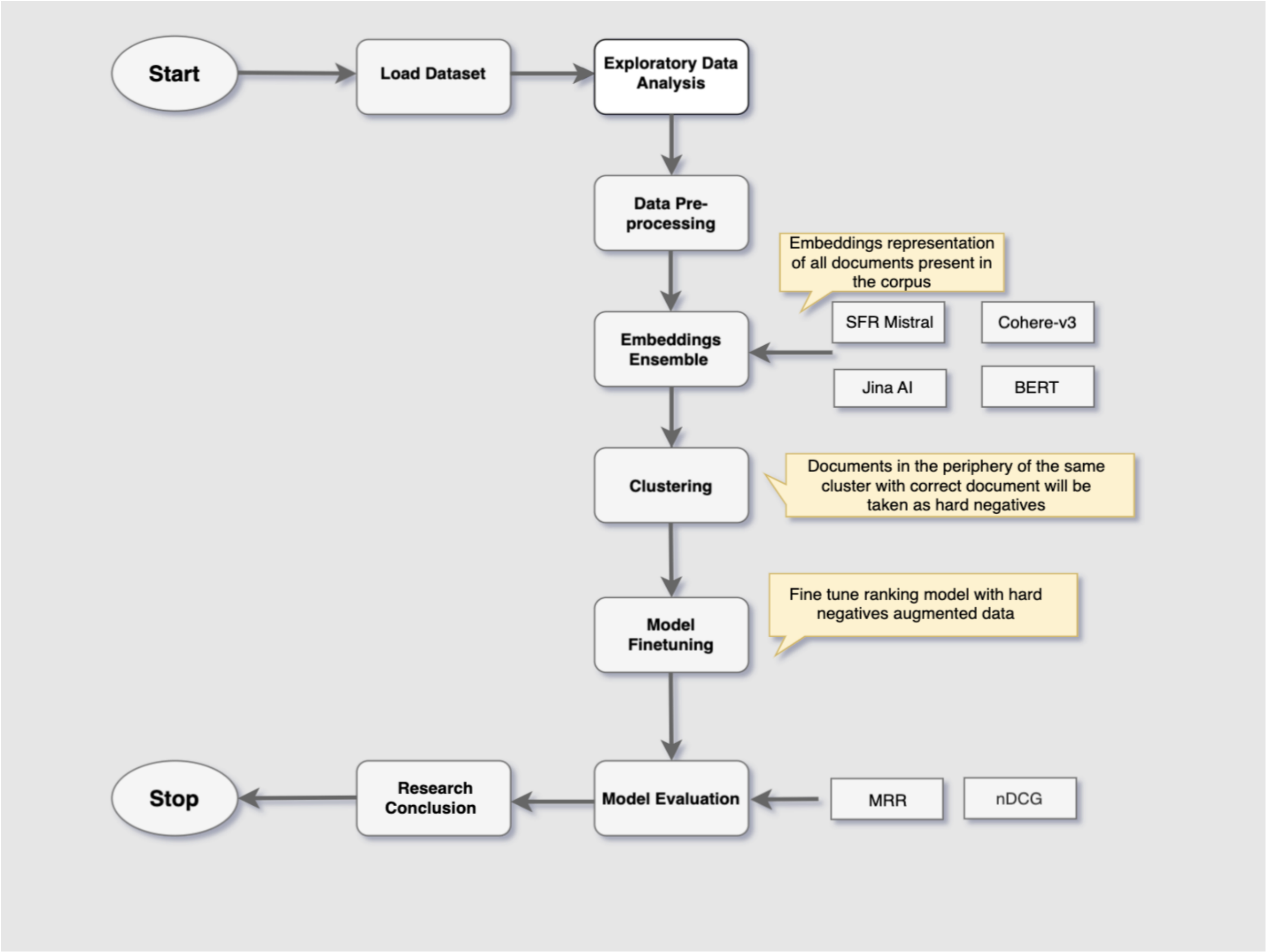

**Figure 1.1** Framework for an ensemble based hard negative selection to enhance retrieval.

## 7.2 Dataset Description and Pre-processing

The dataset selected is private dataset which every organization has for their internal working. This data is related to the specific product/service and mainly utilized by customers for frequently asked question. This type of data mostly resides in company's Confluence, Object Storage, Shared Drive, offline data storage. Dataset consists of 1216 records of query and correct document pairs. Pre-processing of this dataset will require filtering of dataset based on various factors. All documents will undergo initial text normalization procedures, including cleaning, lemmatization, tokenization, and spell correction, among other steps. The pre-processing stage might also entail chunking documents to fit them into the embeddings model, contingent upon the model's token limit and the token count of the document.

## 7.3 Embeddings Ensemble

Embeddings ensembles are a technique used to improve the performance of embedding models by combining multiple embedding models into a single model. This can be done in several ways, such as averaging the outputs of the different models, or using a more sophisticated voting scheme. There are several reasons why embeddings ensembles can be effective. First, different embedding models can capture different aspects of the data. By combining the outputs of multiple models, it is possible to get a more comprehensive representation of the data. In the proposed methodology we propose to use state of the art embeddings models to create ensemble. Some of the embedding models are -

- SFR Mistral Embedding: This model is by Salesforce Research. It is accessible through hugging face transformer library. It is trained on Mistral-7V-v0.1 and E5-mistral-7b-instruct.
- Jina AI's embedding model: This is also available via Hugging Face transformer library. This embedding model supports sequence length of 8192, suitable for dataset which has long documents. The architecture is based on BERT (Devlin et al., 2019b)
- Cohere's embed-v3 model: We can use it via API endpoint, private deployment, AWS SageMaker or Hugging Face transformer library. The sequence length of this model is 512, either documents should be chunked within 512 tokens, or it will be truncated.

## 7.4 Fine-tuning

Fine-tuning cross-encoder models with hard negatives leverages advanced techniques to enhance semantic matching performance in various information retrieval tasks. First, a training dataset is constructed containing query-document pairs with relevance labels. Importantly, besides positive pairs reflecting genuine semantic similarity, "hard negatives" are also included. These are documents demonstrably dissimilar to the query despite potential surface-level similarities, challenging the model to discriminate effectively. During training, the embeddings are compared using a distance metric (e.g., cosine similarity) and fed into a loss function. Fine-tuning involves pre-training a model on a general corpus and then adjusting its parameters while optimizing the loss function specifically on the prepared dataset. Cross encoder models are more powerful than embedding model. It is recommended to fine -tune them and use it on top of embedding result to re-rank documents. Some of the cross- encoder models are –

- Cohere rerank-english-v2.0: This model acts as a text relevance estimator. It takes a query and a collection of texts as input and outputs an ordered list, where each text is annotated with a relevance score calculated based on the provided query. The architecture is based on cross-encoder models (MacAvaney et al., 2019a; Li et al., 2020)
- bge-reranker-base: bge stands for BAAI general embedding. Unlike embedding models, which produce intermediate vector representations, rerankers directly compare questions and documents to output a relevance score. This score, however, can have any value due to the model's training using cross-entropy loss.

These are training often employs triplet loss functions. These losses encourage the model to embed similar pairs closer in the embedding space compared to dissimilar pairs, including the hard negatives. Additionally, margin-based losses can be incorporated, penalizing the model for embedding dissimilar pairs within a specific margin of the positive pair. Fine-tuning cross-encoder models with hard negatives is a powerful technique for improving semantic matching accuracy. By incorporating carefully selected negative examples and employing suitable loss functions, this approach can push the boundaries of information retrieval tasks, enhancing search relevance and efficiency.

## 7.5 Evaluation Metrics

- Regression Objective Function: In this approach, we compute the cosine similarity between the embeddings of the question and the document. The objective function utilized is the mean squared-error loss.
- Triplet Objective Function: This method involves learning from triplets comprising a question q, a positive relevant document p, and a negative irrelevant document n. The triplet loss aims to minimize the distance between q and p while maximizing the distance between q and n. Mathematically, it can be defined as:

$$\boldsymbol{max}(\,||\,\boldsymbol{S_q} - \boldsymbol{S_p}|| - ||\boldsymbol{S_q} - \boldsymbol{S_n}|| + \boldsymbol{1}, \boldsymbol{0})$$

  where $S_x$ represents ensembled document embedding for q/p/n, $||\cdot||$ denotes a distance metric, and a margin of 1 ensures that q is at least 1 unit closer to p than n.
- End To End Metric: For end-to-end performance analysis we will use MRR - Mean Reciprocal Rank is a measure used to evaluate the performance of a system, often in information retrieval tasks like search engines or question answering.

## 8. Requirements Resources

Below are the requirements for conducting the experiment and developing the pipeline:

- Hardware Requirements:
  - Access to a GPU via a cloud platform for fine-tuning the model.
  - Storage space of at least 100 GB to accommodate the text dataset, extracted training data points, the model itself, and its checkpoints.
- Software Requirements
  - For text pre-processing tasks, Spacy, Gensim, and Stanford CoreNLP will be employed.
  - The HuggingFace Transformers library will be heavily utilized for its user-friendly features, facilitating the loading of datasets and integration with the latest state-of-the-art NLP models.
  - Implementation of Siamese class of models will be carried out using PyTorch and PyTorch Lightning frameworks.
  - The development IDE for Python will be Visual Studio Code.
  - Standard libraries such as Sci-kit Learn, Pandas, NumPy, PyTorch text, and other Python libraries will be utilized.

# 9. Research Plan

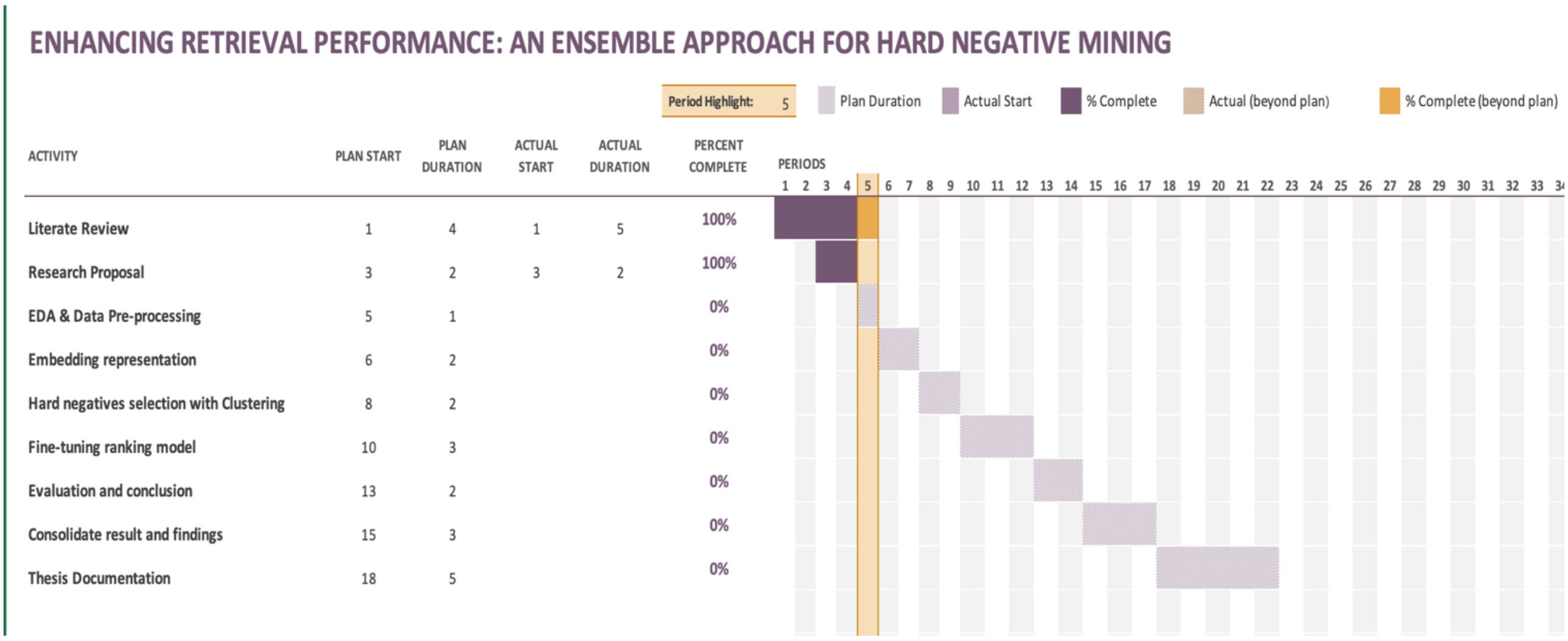

**Figure 2.1** Project Plan.

## 9.1 Risk and Contingency Plan

To anticipate and mitigate potential risks that could affect this project, we've identified the following contingencies:

**Table 1.1** Risk & Contingency plan

| SI | Component of Risk | Contingency measure |
|---|---|---|
| I | Resource and time constraints for incorporating different models for ensemble. | We will replace some of the mentioned models with other available models |
| II | Incompatibility of some part of research methodology with the proposed approach | Substitute the incompatible component with an alternative identified through literature review and implement it accordingly. |

**APPENDIX C: Non-Disclosure Agreement for private dataset**

NDA_signed.pdf

# NON-DISCLOSURE AGREEMENT

**PARTIES**

- This Non-Disclosure Agreement (hereinafter referred to as the **"Agreement"**) is entered into on ___3/01/2024___ (the **"Effective Date"**), by and between Amit Agarwal___, with an address of Principal Applied Scientist at Oracle, India. , (hereinafter referred to as the **"Disclosing Party"**) and Hansa Meghwani___, with an address of Student at LJMU and working professional at Oracle, India__, (hereinafter referred to as the **"Receiving Party"**) (collectively referred to as the **"Parties"**).

**CONFIDENTIAL INFORMATION**

- The Receiving Party agrees not to disclose, copy, clone, or modify any confidential information related to the Disclosing Party and agrees not to use any such information without obtaining consent.
- "Confidential information" refers to any data and/or information that is related to the Disclosing Party, in any form, including, but not limited to, oral or written. Such confidential information includes, but is not limited to, any information related to the business or industry of the Disclosing Party, such as discoveries, processes, techniques, programs, knowledge bases, customer lists, potential customers, business partners, affiliated partners, leads, know- how, or any other services related to the Disclosing Party.

**RETURN OF CONFIDENTIAL INFORMATION**

- The Receiving Party agrees to return all the confidential information to the Disclosing Party upon the termination of this Agreement.

**OWNERSHIP**

- This Agreement is not transferable and may only be transferred by written consent provided by both Parties.

**GOVERNING LAW**

- This Agreement shall be governed by and construed in accordance with the laws of ____LJMU___ .

**SIGNATURE AND DATE**

- The Parties hereby agree to the terms and conditions set forth in this Agreement and such is demonstrated by their signatures below:

DISCLOSING PARTY

Name: Amit Agarwal

Signature:

Date: 3/01/2024

RECEIVING PARTY

Name: Hansa Meghwani

Signature:

Date: 3/01/2024